\documentclass[a4paper,fleqn,usenatbib]{mnras}

\usepackage{newtxtext,newtxmath}

\usepackage[T1]{fontenc}
\usepackage{ae,aecompl}

\usepackage{graphicx}	
\usepackage{amsmath}	
\usepackage{amssymb}	
\usepackage{rotating}
\usepackage{multirow}

\title[Solar system analogues II]{ Formation of Solar system analogues II:\\
  post-gas phase growth and water accretion in extended discs via N-body simulations  
 }                                               

\author[M. P. Ronco et al.]{
M. P. Ronco,$^{1,2,3,4}$\thanks{E-mail: mronco@astro.puc.cl}
G. C. de El\'{\i}a$^{1,2}$\thanks{E-mail: gdeelia@fcaglp.unlp.edu.ar}
\\
$^{1}$Instituto de Astrof\'{\i}sica de La Plata, CCT La Plata - CONICET, UNLP, Paseo del Bosque S/N, (1900) La Plata, Argentina\\
$^{2}$Facultad de Ciencias Astron\'omicas y Geof\'{\i}sicas, Universidad Nacional de La Plata, Paseo del Bosque S/N (1900), La Plata, Argentina\\
$^{3}$Instituto de Astrof\'{\i}sica, Pontificia Universidad Cat\'olica de Chile, Santiago, Chile\\
$^{4}$N\'ucleo Milenio de Formaci\'on Planetaria (NPF), Chile\\
}

\date{Accepted XXX. Received YYY; in original form ZZZ}

\pubyear{2018}

\begin{document}
\label{firstpage}
\pagerange{\pageref{firstpage}--\pageref{lastpage}}
\maketitle


\begin{abstract}
This work is the second part of a project that attempts to analyze the formation of Solar 
system analogues (SSAs) from the gaseous to the post-gas phase, in a self-consistently way. 
In the first paper (PI) we presented our model of planet formation during the gaseous 
phase which provided us with embryo distributions, planetesimal surface density, 
eccentricity and inclination profiles of SSAs, considering different 
planetesimal sizes and type I migration rates at the time the gas dissipates. In this 
second work we focus on the late accretion stage of SSAs using the 
results obtained in PI as initial conditions to carry out N-body simulations. One of 
our interests is to analyze the formation of rocky planets and their final water contents 
within the habitable zone. Our results show that the formation of potentially habitable 
planets (PHPs) seems to be a common process in this kind of scenarios. However, the 
efficiency in forming PHPs is directly related to the size of the planetesimals. The 
smaller the planetesimals, the greater the efficiency in forming PHPs. We also analyze 
the sensitivity of our results to scenarios with type I migration rates and gap-opening 
giants, finding that both phenomena act in a similar way. These effects seem to favor 
the formation of PHPs for small planetesimal scenarios and to be detrimental for scenarios 
formed from big planetesimals. Finally, another interesting result is that the formation 
of water-rich PHPs seems to be more common than the formation of dry PHPs.

\end{abstract}




\section{Introduction}
\label{sec:sec1}

During the past years, improvements in the exoplanet detection techniques and the increase in the quantity and quality of 
the missions, have led to the discovery of exoplanets of all kinds. To date, the number of confirmed exoplanets rises to 3798 
and continues growing \citep[\url{http://exoplanet.eu/}][]{Schneider2011}. In particular, the number of rocky planets 
has grown the most during the last years, discovered mainly by the Kepler mission (\url{https://www.nasa.gov/mission_pages/kepler}). 
The number of gas giant 
exoplanets in orbits with semimajor-axis greater than $\sim$ 1~au has also grown, and many of these rocky and gas giant planets
are part of multiple planetary systems, 633 until now. Moreover, multiple planetary systems with more than one gas giant planet have also 
been detected. From this, many of these multiple planetary systems present a central star similar to our sun. However, to date, we have 
not yet been able to detect planetary systems similar to our own Solar system. This is, planetary systems formed simultaneously by at least 
a giant planet with a mass of the order of Saturn or Jupiter in the outer zones of the disc, and by rocky planets in the inner zone, 
particularly by rocky planets within the habitable zone.

Several works studied the formation of terrestrial planets and water accretion process in different dynamical scenarios or within the 
framework of the formation of our Solar system. For example, \citet{Raymond2004} and \citet{Raymond2006} 
analyzed the effects produced by a Jovian planet located in the outer zones of the disc on the formation and dynamics of terrestrial 
planets around sun-like stars. Later, \citet{Mandell2007} explored the formation of terrestrial-type planets in planetary systems
with one migrating gas giant around solar-type stars. They also considered planetary systems with one inner migrating gas giant and
one outer non-migrating giant planet. \citet{FoggNelson2009} examined the effect of giant planet migration on the formation of inner 
terrestrial planet systems and found that this phenomena does not prevent terrestrial planet formation.
\citet{Zain2017} focused on the formation and water delivery within the habitable zone in different dynamical environments and
also around solar-type stars. These authors considered planetary systems that harbor Jupiter or Saturn analogues around the location of
the snowline and found that the formation of water-rich planets within the habitable zones of those systems seems to be a common process. 
More recently \citet{RaymondIzidoro2017} showed that the inner Solar system's water could be a result of the formation of the giant planets.
These authors find that when the gas disc starts to dissipate, a population of scattered eccentric planetesimals cross the 
terrestrial planet region delivering water to the growing Earth. 

During our first work \citep[][hereafter PI]{Ronco2017} we improved our model of planet formation which calculates
the evolution of a planetary system during the gaseous phase to find, via a population synthesis analysis, suitable scenarios and 
physical parameters of the disc to form Solar system analogues (SSAs from now on). We were specially interested in the planet 
distributions, and in the surface density, eccentricity and inclination profiles for the planetesimal population at the end of the gas-phase
obtained considering different formation scenarios, with different planetesimal sizes and different type I migration rates.

The aim of this second work is to analyze the formation of SSAs during the post-gas phase through the development of N-body simulations,
focusing on the formation of rocky planets and their water contents in the inner regions of the disc, particularly on the habitable zone.
To do so, and in order to analyze the formation of SSAs in a self-consistently way, we make use of the results obtained in PI as the initial
conditions for our N-body simulations and form final planetary systems from different planetesimal sizes. Many works in the literature that analyze
the post-gas phase formation of planetary systems, in particular the formation of rocky planets in the inner regions of the disc, considered
arbitrary initial condition \citep{Chambers2001,Raymond2004,Raymond2006,OBrien2006,Mandell2007,Raymond2009,Walsh2011,Roncodeelia2014}. 
However, as it was already shown by \citet{Ronco2015}, more realistic initial conditions, obtained from a model of planet formation,
lead to different accretion histories of the planets that remain within the habitable zone than considering \emph{ad hoc} initial conditions. 
We also analyze the differences in the results of planetary systems that were affected by the type I migration regime \citep{Tanaka2002}
during the gas-phase, and by planetary systems with gas giant planets that were able to open a gap in the disc and were affected by
type II migration.

This work is then organized as follow. In Section 2, we
summarize our model of planet formation explained in detailed in PI; in Section 3, we describe the
initial conditions for the development of our N-body simulations, obtained from the results of PI; in
Section 4, we describe the most important properties of the N-body code and then show our main results in Section 5. Then, 
we compare our results with the observed potentially habitable planets and with the current
gas giant population in Section 6. Finally, discussions and conclusions are presented
in Section 7 and Section 8, respectively.


\section{Summary of the gas-phase planet formation model}
\label{sec:sec2}
In PI we improved a model of planet formation based on previous works 
\citep{Guilera2010,Guilera2011,Guilera2014}, which we now called P{\scriptsize LANETA}LP. This 
code describes the formation and evolution in time of a planetary system during the gaseous phase,
taking into account many relevant phenomena for the planetary system formation. The model presents a protoplanetary disc, 
which is characterized by a gaseous component, evolving due to an $\alpha$-viscosity driven accretion \citep{Pringle1981} and 
photoevaporation \citep{DangeloMarzari2012} processes, and a solid component represented by a planetesimal and an
embryo population. The planetesimal population is being subject 
to accretion, ejection \citep{Laakso2006} and scattering \citep{IdaLin2004a,Alibert2005} by the embryos, and radial drift due to gas drag including the Epstein, Stokes
and quadratic regimes. The embryo population grows 
by accretion of planetesimals \citep{Inaba2001}, gas \citep{IdaLin2004a,Miguel2011,Mordasini2009,TanigawaIkoma2007}, and due to the fusion between
them taking into account their atmospheres \citep{InamdarSchlichting2015}, while they
migrate via type I \citep{Tanaka2002} or type II \citep{Crida2006} migration mechanisms. We also consider that the evolution of the eccentricities and inclinations
of the planetesimals are due to two principal processes: the embryo gravitational excitation \citep{Ohtsuki2002}, and the damping due to
the nebular gas drag \citep{Rafikov2004,Chambers2008}.
The particularities and the detailed description of the implementation of each phenomena can be found in PI and in \citet{Guilera2014}.

In PI we also described the method, initial parameters and initial formation scenarios considered to synthesize a population of
different planetary systems around a solar-mass star, with the particular aim of forming SSAs. 
Following \citet{Andrews2010}, the gas and solid surface density profiles are modeled as
\begin{align}
\Sigma_{\text{g}}(R) &= \Sigma^{0}_{\text{g}}\left(\dfrac{R}{R_{\text{c}}}\right)^{-\gamma}e^{-\left(\frac{R}{R_{\text{c}}}\right)^{2-\gamma}},\\
\Sigma_{\text{p}}(R) &= \Sigma^{0}_{\text{p}}\eta_{\text{ice}}\left(\dfrac{R}{R_{\text{c}}}\right)^{-\gamma}e^{-\left(\frac{R}{R_{\text{c}}}\right)^{2-\gamma}}
\label{eq:densities}
\end{align}
where  $\Sigma^{0}_{\text{p}}$ and $z_0\Sigma^{0}_{\text{g}}$ are normalization constants, with $z_0 = 0.0153$ the primordial abundance of heavy elements in the sun
\citep{Lodders2009}, and $\eta_{\text{ice}}$ is a parameter that represents the increase in the amount of solid material due to the condensation of water beyond
the snowline, given by 1 if $R \ge R_{\text{ice}}$ and given by $1/\beta$ if $R < R_{\text{ice}}$, where $R_{\text{ice}}$ is the position of the snowline at 2.7~au \citep{Hayashi1981}.

The variable parameters, which were taken randomly 
from range values obtained from the same observational works \citep{Andrews2010}, are the mass of the disc, $M_{\text{d}}$, between $0.01\text{M}_\odot$ and $0.15\text{M}_\odot$,
the characteristic radius $R_{\text{c}}$ between 20~au and 50~au, and the density profile gradient $\gamma$ between
0.5 and 1.5. The factor $\beta$ in the solid surface density profile is also taken randomly between 1.1 and 3. 
The viscosity parameter $\alpha$ for the disc and the rate of EUV ionizing photons $f_{41}$ have a uniform distribution in log between
$10^{-4}$ and $10^{-2}$, and between $10^{-1}$ and $10^4$, respectively \citep{DangeloMarzari2012}.

The formation scenarios considered include different planetesimal sizes ($r_{\text{p}} = 100$~m, 1~km, 10~km and 100~km)
but we consider a single size per simulation. The type I migration rates for isothermal-discs were
slowed down in $1\%$ ($f_{\text{migI}}=0.01$) and $10\%$ ($f_{\text{migI}}=0.1$). We also considered the extreme cases in where type I migration
was not taken into account ($f_{\text{migI}}=0$) or was not slowed down at all ($f_{\text{migI}}=1$).

We also constrained the disc lifetime $\tau$, between 1 Myr and 12 Myr, following \citet{Pfalzner2014}, and considering only those protoplanetary discs
whose combinations between $\alpha$ and $f_{41}$ satisfied this condition.

Finally, the random combinations of all the parameters gave rise to a great diversity of planetary systems which were analyzed in PI. From this
  diversity, only 4.3\% of the total of the developed simulations formed planetary systems that, at the end of the gaseous phase were SSAs. Recalling from
  PI, we define SSAs as those planetary systems that present, at the end of the gaseous phase, only rocky planets in the inner region of the disc and at least one
  gas giant planet, like Jupiter or Saturn, beyond 1.5~au. The great majority of these SSAs  
were found in formation scenarios with small planetesimals and low/null type I migration rates. These SSAs presented different properties 
and architectures, for example, different number of gas giant planets and icy giant planets, and different final planetesimal profiles. 
Moreover, from a range of disc masses between $0.01M_\odot$ and $0.15M_\odot$, SSAs were formed only within discs more massive than
$M_{\text{{d}}} \ge 0.04\text{M}_\oplus$ since low-mass discs have not enough solid 
material to form massive cores before the disc dissipation timescales. However, these SSAs were not sensitive to the different values of 
$R_{\text{c}}$, $\gamma$, $\beta$, $\alpha$ and $\tau$, since we found them throughout the whole range of values 
considered for each of these parameters. The mean values for the disc parameters that formed SSAs were $<M_{\text{d}}>=0.1M_\odot$, $<R_{\text{c}}>=32.8$~au, $<\gamma>=0.95$,
$<\beta>=2.07$, $<\alpha=3.4 \times 10^{-3}$ and $<\tau>=6.45$~Myr.

It is important to remark that the planetary systems obtained at the end of the gaseous phase and which are going to be consider as initial
  conditions for the post-gas growth, were formed within the framework of a planetary formation code that, although presents important physical
  phenomena for the formation of a planetary system during the gaseous phase, it also presents certain limitations. Some of these limitations are,
  for example, the use of a radial 1D model to compute the evolution of a classical $\alpha$-disc, the consideration of isothermal discs and the
  corresponding type I migration rates for such discs, and the lack of the calculation of the gravitational interactions between embryos. Then,
  the initial conditions for the N-body simulations carried out in this work are dependent on the physics of our planetary formation model and
  more detailed treatments of the physics could affect the final embryo and planetesimal configurations of planetary
  systems at the end of the gaseous phase. For example, recent works which consider non-isothermal discs have shown that type I migration rates
  could be very different with respect to previous isothermal type I migration rates \citep{Paardekooper2010,Paardekooper2011}. Moreover, for non-isothermal
  discs, migration convergent zones are generated, acting as traps for migrating planets \citep{HellaryNelson2012,Bistch2015}. On the other hand, the presence
  of a dead zone along the disc could also affect both, the planet migration and the planetesimal radial drifts. \citet{GuileraSandor2017}
  showed that even in isothermal discs, the presence of a dead zone in the disc leads to pressure maxima at the edges of such zones acting as traps for
  migrating planets and also as convergent zones for drifting planetesimals. Finally, alternative models for $\alpha$-discs, as wind-discs, have been proposed
  to model the evolution of the gaseous component of protoplanetary discs, showing different final disc structures \citep{Suzuki2010,Suzuki2016}. These differences
  are important and could affect  models of planetesimal formation and planet migration.

\section{Initial conditions}
\label{sec:sec3}

The goal of this second work is to study and analyze the evolution of the post-gas phase of planetary systems that resulted
in SSAs at the end of the gaseous phase. From this,
Fig. 8 of PI showed some of the embryo and planetesimal distributions of the different scenarios of SSAs at the end of the gaseous phase.

Some of the final configurations obtained after the gas is completely dissipated are used in this work as initial conditions
to develop N-body simulations.
Since one of our main goals is to analyze the formation of rocky planets within the habitable zone (HZ) of SSAs, we focuse on the long-term
evolution of planetary systems between 0.5~au and 30~au. We are not interested in the formation of close-in planets very near the central star.
Thus, we do not consider planetary embryos neither planetesimals inside 0.5~au.

\begin{figure*}
  \includegraphics[angle=270, width=0.99\textwidth]{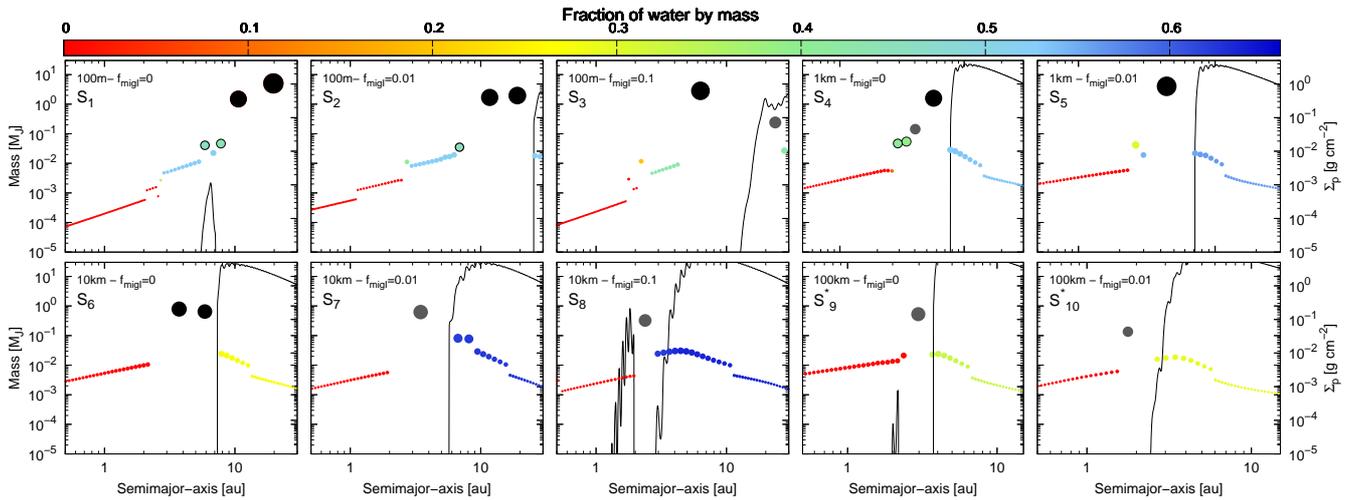}
  \caption{Chosen SSAs configurations at the end of the gaseous phase, to develop N-body simulations in a 
    mass vs. semimajor-axis vs planetesimal surface density plane. Each colored point represents a planetary embryo, the black and
    grey big dots represent Jupiter and Saturn analogues, respectively, while the colored points with black borders represent
    icy giants, like Neptune-like planets. 
    The solid black line in all the planetary systems represents the planetesimal density profile when the gas is already gone.
    The colorscale represents the final amounts of water by mass in each planet and the size of each planet is
    represented in logarithmic scale.}
  \label{fig:fig1}
\end{figure*}

Table \ref{tab:tab1} and Fig. \ref{fig:fig1} show the main properties of the chosen outcomes of the results of PI that are going to be used
to analyze the late-accretion stage. Due to the high computational cost of N-body simulations of this kind it is not viable to develop
  simulations for each SSA found. Thus we have to chose the most representative one for each formation scenario. Therefore, the chosen outcomes shown in
  Fig. \ref{fig:fig1}, that represent the embryo distributions and the planetesimal surface density profiles of SSAs at the
  end of the gaseous phase, were carefully selected from each formation scenario, this is taking into account the different planetesimal sizes and
  different type I migration rates, considering that it was the most representative of the whole group.
  
\begin{table*}
  \centering
  \caption{Chosen configurations of SSAs at the end of the gas-stage to be used as initial conditions for the development of N-body 
    simulations. The total mass in embryos (without considering the mass of the Jupiter or Saturn analogues) and planetesimals is calculated
    between 0.5~au and 30~au except for the $*$ cases, which represent scenarios that had been cutted of until 15 au in order to use 1000 planetesimals. 
    The mass in embryos inside and outside, represent the mass in embryos inside and outside the giants location. The last four columns 
    of this table show 
     characteristics concerning water contents. $\beta$ is a variable factor that represents the jump in the solid surface density due 
    to the water condensation beyond the snowline. The maximum water content by mass is represented in \%.}
  \label{tab:tab1}
  \begin{tabular}{l|c|c|c|c|c|c|c|c|c|c} 
    \hline
    &\multicolumn{2}{c}{Formation scenario}&\multicolumn{2}{c}{Total mass in}&\multicolumn{2}{c}{Mass in embryos}&\multicolumn{4}{c}{Initial water contents}\\
    \hline
    Scenario & Planetesimal & $f_{\text{migI}}$ & Embryos & Planetesimals & Inside & Outside & $\beta$ & Max. water & Dry & Water-rich \\
    & size & & & & & & & contents \% & embryos & embryos \\
    \hline
    $S_1$ & 100~m & 0 & $66.65\text{M}_\oplus$ & $2.5\times10^{-4}\text{M}_\oplus$ & $66.65\text{M}_\oplus$ & $0\text{M}_\oplus$ & 2.12 & 52.8 & 72 & 14 \\ 
    $S_2$ & 100~m & 0.01 & $86.33\text{M}_\oplus$ & $8.28\text{M}_\oplus$ & $75.17\text{M}_\oplus$ & $11.16_\oplus$ & 2.11 & 52.6 & 49 & 15 \\ 
    $S_3$ & 100~m & 0.1 & $35.13\text{M}_\oplus$ & $13.78\text{M}_\oplus$ & $26.54\text{M}_\oplus$ & $8.59\text{M}_\oplus$ & 1.75 & 42.8 & 62 & 10 \\ 
    $S_4$ & 1~km & 0 & $114.36\text{M}_\oplus$ & $174.50\text{M}_\oplus$ & $61.47\text{M}_\oplus$ & $52.89\text{M}_\oplus$ & 2.14 & 53.3 & 28 & 25 \\
    $S_5$ & 1~km & 0.01 & $93.61\text{M}_\oplus$ & $156.55\text{M}_\oplus$ & $48.47\text{M}_\oplus$ & $45.15\text{M}_\oplus$ & 2.27 & 55.9 & 25 & 28 \\
    $S_6$ & 10~km & 0 & $89.02\text{M}_\oplus$ & $195.30\text{M}_\oplus$ & $42.59\text{M}_\oplus$ & $46.43\text{M}_\oplus$ & 1.37 & 27.0 & 23 & 23 \\
    $S_7$ & 10~km & 0.01 & $120.91\text{M}_\oplus$ & $226.43\text{M}_\oplus$ & $23.41\text{M}_\oplus$ & $97.5\text{M}_\oplus$ & 2.66 & 62.4 & 23 & 20 \\
    $S_8$ & 10~km & 0.1 & $134.66\text{M}_\oplus$ & $271.84\text{M}_\oplus$ & $18.87\text{M}_\oplus$ & $115.79\text{M}_\oplus$ & 2.75 & 63.6 & 23 & 34 \\
    $S^*_9$ & 100~km & 0 & $116.44\text{M}_\oplus$ & $164.40\text{M}_\oplus$ & $66.22\text{M}_\oplus$ & $50.22\text{M}_\oplus$ & 1.49 & 32.8 & 22 & 24 \\
    $S^*_{10}$ & 100~km & 0.01 & $68.88\text{M}_\oplus$ & $141.88\text{M}_\oplus$ & $21.96\text{M}_\oplus$ & $46.92\text{M}_\oplus$ & 1.41 & 29.1 & 19 & 28  \\
    \hline
  \end{tabular}
\end{table*}

As it was explained in PI, a water radial distribution for the embryos and planetesimals was included in our model of planet formation.
The colorscale in Fig. \ref{fig:fig1} shows the water contents in embryos at the end of the gaseous phase.
As the fraction per unit mass of water in embryos and planetesimals, at the beginning of the gaseous phase, was given by
\begin{equation}
  f_{\text{H}_2\text{O}}(R)=
  \begin{cases}
    0,~\text{if}~R \le R_{\text{ice}}, \\
    \\
    \dfrac{\beta - 1}{\beta},~\text{if}~R > R_{\text{ice}},
  \end{cases}
  \label{eq:water-distribution}
\end{equation}
the water contents in embryos at the end of the gaseous phase, which represent the initial water contents for the post-gas 
phase evolution, vary, on the one hand according to the $\beta$ value considered for each scenario, and on the other hand due to 
the growth of each embryo during the gas phase. As we mentioned before, $\beta$ takes random values from 1.1 ( which represents a 9\% of water by mass) to
3 (which represents a 66\% of water by mass). The planetesimal population remains always the same regarding water contents: planetesimals
inside 2.7~au remain dry and planetesimals beyond 2.7~au contain the amount of water by mass given by Eq. \ref{eq:water-distribution}.
As we already mentioned, the last four columns of table \ref{tab:tab1} show the main properties related to water contents in each formation scenario
considered to develop N-body simulations.

It is important to note that, at the end of the gaseous phase, we are able to distinguish which is the origin of the water content of each planet.
This is, we can know the percentage of water by mass at the end of the gaseous phase due to the accretion of planetesimals
and the percentage of water by mass due to the fusion between embryos.

An interesting difference between some of the formation scenarios considered, is that due to the position
of the formed giant planets during the gaseous phase, not all the SSAs present a water-rich embryo population
inside the giant locations. This topic will play an important role in the analysis of the water contents in planets within the HZ 
at the end of the N-body simulations, and it will be attended in the following sections. 

\section{Properties of the N-body simulations}
\label{sec:sec4}

Our N-body simulations begin when the gas is already dissipated in the protoplanetary disc. The numerical code used to 
carry out our N-body simulations is the MERCURY code developed by \citet{Chambers1999}. In particular, we adopted the 
hybrid integrator, which uses a second-order mixed variable symplectic algorithm for the treatment of the interaction between
objects with separations greater than 3 Hill radii, and a Burlisch-Stoer method for resolving closer encounters.

The MERCURY code calculates the orbits of gas giants, planetary embryos and planetesimals. When these orbits cross each other, 
close encounters and collisions between the bodies may occur and are registered by the MERCURY code.
It is important to mention that in our simulations, collisions between all the bodies
are treated as inelastic mergers, which conserve mass and water content and because of that our model does not consider water loss
during impacts. As a consequence, the final mass and water contents of the resulting planets should be considered as upper limits.
 
Given the high numerical cost of the N-body simulations and in order to perform them in a viable CPU time, we only 
consider gravitational between planets (embryos and giant planets), and between planets and planetesimals, but planetesimals are not self-interacting.
This is, planetesimals perturb and interact with embryos and giant planets but they ignore one another completely and they cannot collide with each other \citep{Raymond2006}.
It is important to remark that these planetesimals are in fact \emph{superplanetesimals}.While the size of the planetesimals is only relevant during the gas phase
  when planetesimals interact with the gaseous component of the disc due to the drift of the nebular gas (see PI), the size ceases to be relevant in the post-gas phase during the N-body
  simulations. Thus, due to the numerical limitations of N-body simulations regarding the limited number of bodies that has to be used, it is not possible to
consider, in a consistently way, realistic planetesimals of the proposed sizes in PI, this is planetesimals of 100~m, 1~km, 10~km and 100~km. Instead,
we have to consider higher mass \emph{superplanetesimals} that represent an idealized swarm of a larger number of real size planetesimals.
Thus, from now on when we speak about planetesimals we are referring to \emph{superplanetesimals}.

As we already mentioned, we are interested in the post-gas evolution of planetary systems similar to our own, and this is why
we consider planetary embryos and planetesimals, in extended discs, between 0.5~au and 30~au. The number of embryos varies between 
40 and 90 according to each scenario and present physical densities of 3~gr~cm$^{-3}$. We consider 1000 planetesimals in all the developed simulations
with physical densities of 1.5~gr~cm$^{-3}$. Since the remaining mass
in planetesimals, at the end of the gaseous phase, in simulations with planetesimals of 100~km and type I migration rates reduced 
to a 0.1\% are too high, we cut the disc until 15~au in order to be able to consider 1000 planetesimals. The choice of considering 
only 1000 planetesimals, all with the same mass for each scenario, has to do with the fact that considering more planetesimals would 
be unviable taking into account the time needed to run these simulations, and since this is a number that will allow us to study in a 
moderately detailed way the evolution of this population throughout the whole disc.

Since terrestrial planets in our Solar system may have formed within 100~Myr - 200~Myr \citep{Touboul2007,Dauphas2011,Jacobson2014}, 
we integrate each simulation for 200 Myr. To compute the inner orbit with enough precision, we use a maximum time step of 
six days which is shorter than 1/20~th of the orbital period of the innermost body in all the simulations. The reliability of all our
simulations is evaluated taking into account the criterion of energy $|dE/E| < E_{\text{t}}$, where $E_{\text{t}}$ is an energy threshold 
and it is considered as $10^{-3}$. However, in order to conserve the energy to at least one part in $10^{3}$, we carry out simulations
assuming shorter time steps of four and up to two days, depending on the requirement of each simulation. Those simulations that do not meet this 
energy threshold value are directly discarded.

To avoid any numerical error for small-perihelion orbits, we use a non-realistic size for the sun radius of 0.1~au \citep{Raymond2009}.
The ejection distance considered in all our simulations was 1000~au. This value is realistic enough for the study of extended planetary
systems and is also a value that allows us to reduce the total number of bodies of our simulations as time passes. 
\citet{HiguchiKokubo2007} found that planetesimals with semimajor-axis $> 1000$~au increase their perihelion distances outside the
planetary region (100~au) due to the fact that they would already be affected by galactic tides. In addition, \citet{VerasEvans2013} showed
that wide-orbit planets with semimajor-axis greater than 1000~au are substantially affected by galactic tides while closer-in planets, at 
$\sim$ 100~au, are only affected if the host star is near the galactic center, within the inner hundred parsecs.

\begin{figure}
  \includegraphics[angle=270, width=\columnwidth]{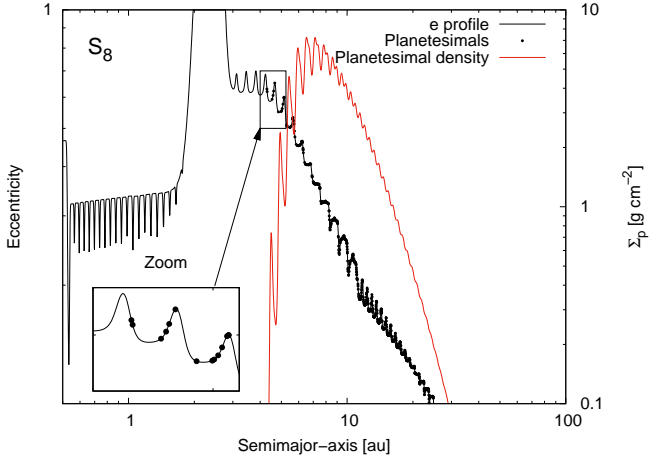}
  \caption{Eccentricity profile (black curve) for the planetesimal surface density profile (red curve) of the end of the gas-phase evolution of
    scenario $S_8$ of Fig. \ref{fig:fig1}. The oscillations in the eccentricity profiles of planetesimals are caused, during the gaseous phase (see PI),
      by the embryo gravitational excitation \citep{Ohtsuki2002} and by the damping due to the nebular gas drag \citep{Rafikov2004,Chambers2008}.
    The zoom shows some planetesimals in black points. All of them, the 1000 planetesimals,
    represent the mass in planetesimals available between 0.5~au and 30~au, located on top of the eccentricity profile.} 
  \label{fig:fig2}
\end{figure}

The semimayor-axes of the planetesimals are generated using an acceptance-rejection method taking into account the surface density profile
of planetesimals shown with black curves in each formation scenario of Fig. \ref{fig:fig1}. As it was explained before, the orbital eccentricities
and inclinations for the planetesimal population are results of the gas-phase evolution (see PI). Each planetesimal surface density profile has an 
eccentricity and inclination profile that are also used as initial conditions for this population. Figure \ref{fig:fig2} shows
the eccentricity profile (black curve) of the planetesimal surface density profile (red curve) of the results of $S_8$, as an example.
For embryos and giant planets, the initial eccentricities and inclinations are taken randomly considering values lower
  than $0.02$ and $0.5^{\circ}$, respectively. Recalling from PI, embryos and giant planets evolve in circular and coplanar orbits during the gaseous phase. Thus,
  to be consistent with our model of planet formation we initiate the N-body simulations with low values for the eccentricities and inclinations. It is woth
  mentioning that the model developed in PI does not yet consider the gravitational self-stirring of embryos and giant planets. Therefore, this limitation in our
  model could change the configurations of our SSAs at the end of the gaseous stage.

The rest of the orbital parameters, 
such as the argument of pericenter $\omega$, longitude of ascending node $\Omega$, and the mean anomaly $M$ are also taken randomly 
between $0^{\circ}$ and $360^{\circ}$ for the gas giants, the embryos and the planetesimals.

Due to the stochastic nature of the accretion process, we perform 10 simulations with different random number seeds per each 
formation scenario (see Table \ref{tab:tab1}). 

\section{Results}
\label{sec:sec5}
In this section we describe the main results of our N-body simulations for the formation of planetary systems from SSAs configurations.
It is worth remembering that the considered architecture of a SSA consists on a planetary system with rocky planets in the inner region
of the disc and at least one giant planet, a Jupiter or Saturn analogue, beyond 1.5~au. The main goal of our work is to study the
formation process of the whole system, focusing on the formation of rocky planets and the water delivery within the HZ.
As we have developed a great number of N-body simulations taking into account different planetesimal sizes and different migration rates, and in
order to show our results in an organized way, we first describe the results of scenarios formed from planetesimals of 100~km, 10~km, 1~km and 100~m
  that did not undergo type I migration and that present gas giants that had not managed to open a gap during the gas phase, as the reference scenarios.

The sensitivity of the results to different type I migration rates and to giant planets that open a gap in the gas disc, are going to be
exposed in the subsequent sections.

\subsection{The habitable zone}
\label{subsec:subsec1}
One of the goals of this work is to analyze the formation of rocky planets within the HZ and their final 
water contents. The HZ was first defined by \citet{Kasting1993} as the circumstellar region inside where a planet 
can retain liquid water on its surface. For a solar-mass star this region remains between 0.8~au and 1.5~au 
\citep{Kasting1993,Selsis2007}. Later, and particularly for the Solar system, \citet{Kopparapu2013,Kopparapuerrata2013} proposed 
that an optimistic HZ is set between 0.75~au and 1.77~au. These limits were estimated based on the belief that Venus
has not had liquid water on its surface for the last 1 billion years, and that Mars was warm enough to maintain liquid
water on its surface. The authors also defined a conservative HZ determined by loss of water and by the maximum greenhouse
effect provided by a CO$_2$ atmosphere, between 0.99~au and 1.67~au.
However, locating a planet within these regions does not guarantee the potential habitability, taking into
account that if the orbits of planets found inside these regions are too eccentric, they could be most of the time outside them,
and this could avoid long times of permanent liquid water on their surfaces. In order to prevent this situation, in previous 
works that also analyze the formation of rocky planets within the HZ, we considered that
a planet is in the HZ and can hold liquid water if it has a perihelion $q \ge 0.8$~au and a aphelion $Q \le 1.5$~au 
\citep{Roncodeelia2014}, or has conservative or optimistic chances to be habitable if it has a perihelion
$q \ge 0.75$~au and a aphelion $Q \le 1.77$~au, and if it has a perihelion $q \ge 0.99$~au and a aphelion $Q \le 1.67$~au, 
respectively \citep{Ronco2015}.

\citet{WilliamsPollard2002} showed that the planetary habitability would not be compromised
if the planet does not stay within the HZ during all the integration. If an ocean is present to act as a heat capacitor, or the atmosphere
protects the planet from seasonal changes, it is primarily the time-averaged flux that affects the habitability over eccentric orbits.
In this sense, \citet{WilliamsPollard2002} determined that the mean flux approximation could be valid for all eccentricities.
However, \citet{Bolmont2016} explored the limits of the mean flux approximation varying the luminosity of the central star and the 
eccentricity of the planet, and found that planets orbiting a $1\text{L}_\odot$ star with eccentricities higher than 0.6 would not be able 
to maintain liquid water throughout the orbit.

\citet{Kopparapu2014} improved the analysis of the HZ boundaries founding a dependence between the limits of the HZ with the masses 
of the planets. Assuming H$_2$O (inner HZ) and CO$_2$ (outer HZ) atmospheres, their results indicated that larger planets have wider
HZs than smaller ones. Particularly, \citet{Kopparapu2014} provided with parametric equations to calculate the corresponding HZ
boundaries of planets of $0.1\text{M}_\oplus$, $1\text{M}_\oplus$ and $5\text{M}_\oplus$. They determined that for a $1\text{L}_\odot$, the inner conservative boundary of a
planet of $0.1\text{M}_\oplus$ is 1.005~au, for a planet of $1\text{M}_\oplus$ is 0.95~au, and for a planet of  $5\text{M}_\oplus$ is 0.917~au, while the
outer limits remain the same for the three of them and is 1.67~au. 

In this work, on the one hand we consider that those planets with masses lower than $0.1\text{M}_\oplus$ present the same inner HZ boundary of 
a planet of $0.1\text{M}_\oplus$, planets with masses between $0.1\text{M}_\oplus$ and $1\text{M}_\oplus$ present the same inner HZ boundaries of 
a planet of $1\text{M}_\oplus$, and planets with masses between $1\text{M}_\oplus$ and $5\text{M}_\oplus$ present the same inner HZ boundaries of 
a planet of $5\text{M}_\oplus$. Planets with masses greater than $5\text{M}_\oplus$ also present the inner HZ boundaries of a planet of $5\text{M}_\oplus$. 

On the other hand, we also calculate the time-averaged incident stellar flux for the inner and outer HZ limits of an eccentric planet, 
in terms of the solar flux following
\begin{equation}
  F^{*}_{\text{eff}} = \dfrac{F_{\text{eff}}}{(1-e^{2})^{1/2}},
  \label{eq:flux}
\end{equation}
where $F_{\text{eff}}$ is the effective flux from a circular orbit. Then, following \citet{WilliamsPollard2002} we can draw constant 
flux curves in the semimajor-axis vs eccentricity plane for the HZ boundaries.

Therefore, given the previous mentioned works and our considerations, the adopted criteria for the definition of the HZ and the
class of potentially habitable planets, PHPs from now on, is as follows:

\begin{enumerate}
\item If a planet presents, at the end of the N-body simulations, a semimajor-axis between its HZ boundaries, and
has a perihelion $q$ greater, and a aphelion $Q$ lower than the limit values for their HZ boundaries, we consider this planet is a PHP class A.
\item If a planet presents a $(a,e)$ pair that remains inside the constant flux curves associated to the inner and outer HZ edges,
and has an eccentricity lower or equal than 0.6, we consider this planet is a PHP class B.
\item Finally, if a planet remains within a 10\% of the inner or the outer constant flux curve, and has an
eccentricity lower or equal than 0.6, we consider this planet is a PHP class C.
\end{enumerate}

\begin{figure}
  \center
  \includegraphics[angle=0, width=1\columnwidth]{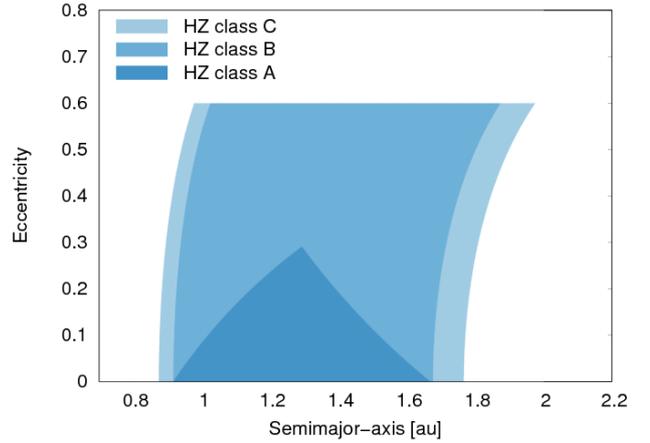}
  \caption{Habitable zone class A, B and C for a planet of $5\text{M}_\oplus$.}
  \label{fig:fig3}
\end{figure}

Although planets of the three classes will be considered as PHPs, the most important ones, taking into account the inner and
outer HZ boundaries and the mean flux definition are those of class A and B. Planets of class C, for being in a region that extends in a 10\%
on both sides of each boundary, are considered as marginally habitable but PHPs at last. Figure \ref{fig:fig3} shows, as an example,
the different HZ regions for a planet of $5\text{M}_\oplus$. These regions are reduced if we consider
less massive planets since larger planets have wider HZs than smaller ones.

Regarding final water contents, we consider that to be a PHP of real interest, a planet has to harbor at least some water content, better if this content is
  of the order of the estimations of water on Earth. The mass of water on the Earth surface is around $2.8\times10^{-4}\text{M}_\oplus$, which is defined as 1 Earth ocean. However, as the actual mass of
water in the mantle is still uncertain, the total mass of water on the Earth is not yet very well known. However, \citet{Lecuyer1998} estimated
that the mantle water is between $0.8\times10^{-4}\text{M}_\oplus$ and $8\times10^{-4}\text{M}_\oplus$, and \citet{Marty2012} estimated a value of $2\times10^{-3}\text{M}_\oplus$.
Thus, taking into account the water reservoirs that should be in the Earth's mantle and the mass of water on the surface, the present-day water content 
on Earth should be around 0.1\% to 0.2\% by mass, which represents between 3.6 to 7.1 Earth oceans. Moreover, \citet{Abe-2000} suggested that the primitive mantle
of the Earth could have contained between 10 and 50 Earth oceans. However, other authors assumed that the current content is not very different from the one
that was accreted during the formation process \citep[see][and references therein]{Raymond2004}. Althought to be a PHP of real interest the planet has to present
  at least some water content, we do not consider a minimum for this value since we could not ensure the potential non-habitability for lower values.

\subsection{Description of the reference scenarios}
\label{subsec:subsec2}

Here we describe the most important characteristics of the four reference scenarios: planetary systems formed from planetesimals of 100~km, 10~km, 1~km and 100~m
and without suffering migration regimes during the gaseous phase.

\subsubsection{Scenarios with planetesimals of 100~km}

\indent{\bf Main perturber:} A gas giant planet similar to a 
Saturn-like\footnote{From PI we remember that our consideration of a Saturn-like planet is a giant planet with an envelope 
that is greater than the mass of the core but its total mass is less than $200\text{M}_\oplus$.} planet acts as the main perturber 
of the systems. This giant is initially located at 2.97~au and then migrates inward due to an embryo-driven migration until it ends up located between 1.57~au and 1.97~au
at the end of the evolution, with a mass ranging between $171.4\text{M}_\oplus$ and $190\text{M}_\oplus$ (this is $1.8\text{M}_{\text{S}}$ and $2\text{M}_{\text{S}}$)
taking into account the whole group of simulations.

\begin{figure*}
  \includegraphics[angle=270, width=\textwidth]{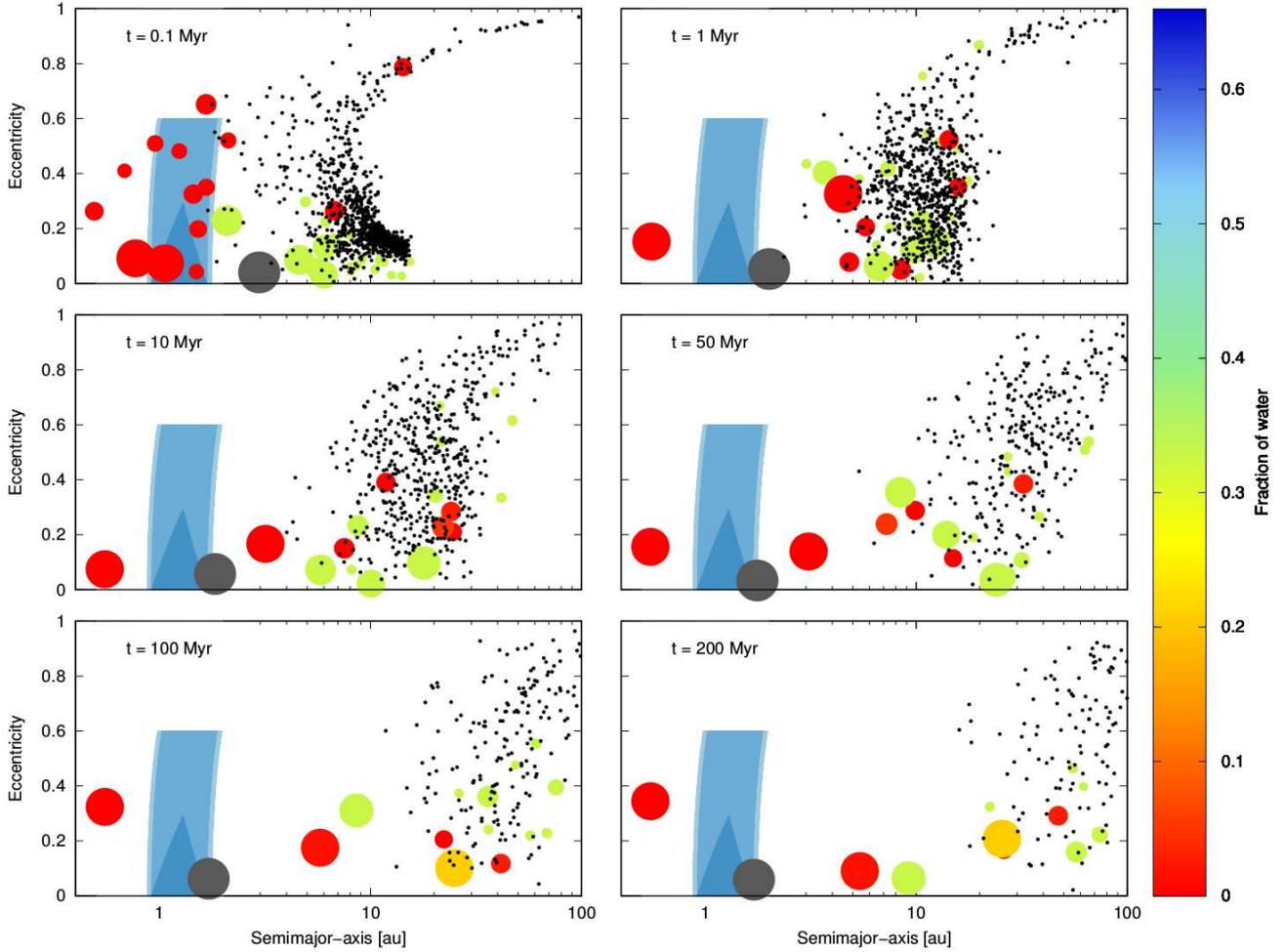}
  \caption{Time evolution of SIM$_4$ of the scenario with planetesimals of 100~km and without considering type I migration.
    The shaded areas represent the HZ regions already described in Section \ref{sec:sec1}. From now on, all the HZ regions plotted
    are the widest possible, which correspond to a planet of $5\text{M}_\oplus$. Planets are plotted as colored circles,
    planetesimals are plotted with black dots and the giant planet, which is a Saturn-like planet in this case, is represented in grey.
    The size of the Saturn-like planet is fixed and is not in the same scale for the rest of the planets. The same happens with the
      planetesimal population, each planetesimal adopts a size of half the size of the smallest embryo, independently of its mass. Then, the size
      for the rest of the planets is represented by a lineal relation between the mass and the size, where the size for the smallest embryo is the double
      of the size of the planetesimals, and the size of the biggest planet is smaller that the size for the Saturn-like planet.    
    The color scale represents the fraction of water of the planets with respect to their total masses. This scenario always presents a Mega-Earth in the inner region per simulation.
    Half of them are completely dry as it can be seen in this case. This scenario does not form PHPs within the HZ.}
  \label{fig:fig4}
\end{figure*}

{\bf General dynamical evolution:} The dynamical evolution of all the simulations performed with planetesimals of 100~km is very similar. Figure \ref{fig:fig4} six snapshots in time of the semimajor-axis eccentricity plane of SIM$_{4}$ as an example of the whole group. The initial conditions correspond to $S^*_{9}$ in Fig. \ref{fig:fig1}.
From the very beginning, the inner embryo population, which does not coexist with planetesimals, gets high 
eccentricities due to their own mutual gravitational perturbations and particularly with the giant planet. The planetary embryos evolved
through collisions with planetesimals and other embryos. Particularly, the collisional events between the inner embryos happened within the 
first million year of evolution where, at that time, only one planet remains in the inner zone of the disc.

{\bf Mass remotion:} Between a 30\% and a 40\% of the initial embryos, and between a 6\% and a 14\% of the initial planetesimals survive at the end of the simulations.
Considering mean values for all the simulations, a 2.76\% of the initial embryos collide with the central star while a 24.7\% is accreted, and a 39\% of
is ejected from the systems. The rest of them remains in the inner zone\footnote{From now on, when we talk 
about planets or planetesimals from the inner zone, we are referring to those planets or planetesimals that are located inside the giants 
location at the beginning of the N-body simulations. Similarly, planets or planetesimals from the outer zone are 
those planets or planetesimals that are located beyond the giants location at the beginning of the N-body simulations.} as a Mega-Earth\footnote{We classify the formed rocky planets as Mega-Earths if
they present masses between $8\text{M}_\oplus$ and $30\text{M}_\oplus$, Super-Earths if they present masses between $2\text{M}_\oplus$ and $8\text{M}_\oplus$
and Earth-like planets if their masses are lower than $2\text{M}_\oplus$ and greater than half an Earth.} and in the outer zones of the disc coexisting 
simultaneously with a planetesimal population. 
The planetesimals suffer a similar process. They are located behind the giant and are perturbed by the giant and the surrounding 
planets. Considering all the simulations, the great majority ($\sim$ 84.3\%) are ejected from the systems and only a few 
collide with the central star (3.1\%) or are accreted (2.5\%) by the surrounding planets. Following these results, the accretion of 
planetesimals is an inefficient process in this scenario. The rest of the planetesimals ($\sim 10\%$) presents high eccentricities at the end of
the evolution suggesting that if we extended the simulations by a few million years more, this population would probably end up being completely ejected from the systems. 

{\bf Inner region:}
Inside the final location of the giant we always find a Mega-Earth with final mass in a range of
$14.5\text{M}_\oplus-26\text{M}_\oplus$, located between 0.37~au and 0.79~au. The seeds of these Mega-Earths where initially located always inside the snowline between 0.56~au and 2.37~au, with an
initial mass between $1.70M_\oplus$ and $6.76\text{M}_\oplus$. During the evolution of the systems, these seeds also migrated inwards finally settling between 0.37~au and 0.79~au. Half of these
Mega-Earths are completely dry at the end of the evolution. The other half present amounts of water by mass due to the acretion of very few water-rich planetesimals.
However, none of them lie within the HZ.
This is why this scenario is not interesting from an astrobiological point of view, the efficiency in forming PHPs is completely null.

{\bf Outer region:} Beyond the giant, a mixture of dry and water-rich planets, and a planetesimal 
population with high mean eccentricities can be found until $\sim$ 400~au. This mix of material was found in all the developed simulations and the 
average number of dry planets found in the outermost regions is $\sim$ 4. The gravitational interactions between the gas giant and the 
embryo population cause that some of the internal embryos are injected into the outer zones of the disc, however no water-rich embryos from beyond the snowline are injected in the inner regions.

Finally, after 200 Myr of evolution, the planetary systems resulting from $S^*_{9}$ in figure \ref{fig:fig1} are composed by a dry Mega-Earth at around 0.55~au in half of the cases, a Saturn-like planet near or in the outer edge of the HZ, and of several planets beyond the giant planet.

\subsubsection{Scenarios with planetesimals of 10~km}

{\bf Main perturbers:} In these systems two Jupiter-like planets with masses of $0.79M_{\text{J}}$ and $0.64M_{\text{J}}$, and initially located at 3.73~au and 5.88~au, respectively, act as the main perturbers. After 200 Myr of evolution the innermost giant is located at around 2.85~au and the outermost giant is located at around 5.78~au.

\begin{figure*}
  \includegraphics[angle=270, width=\textwidth]{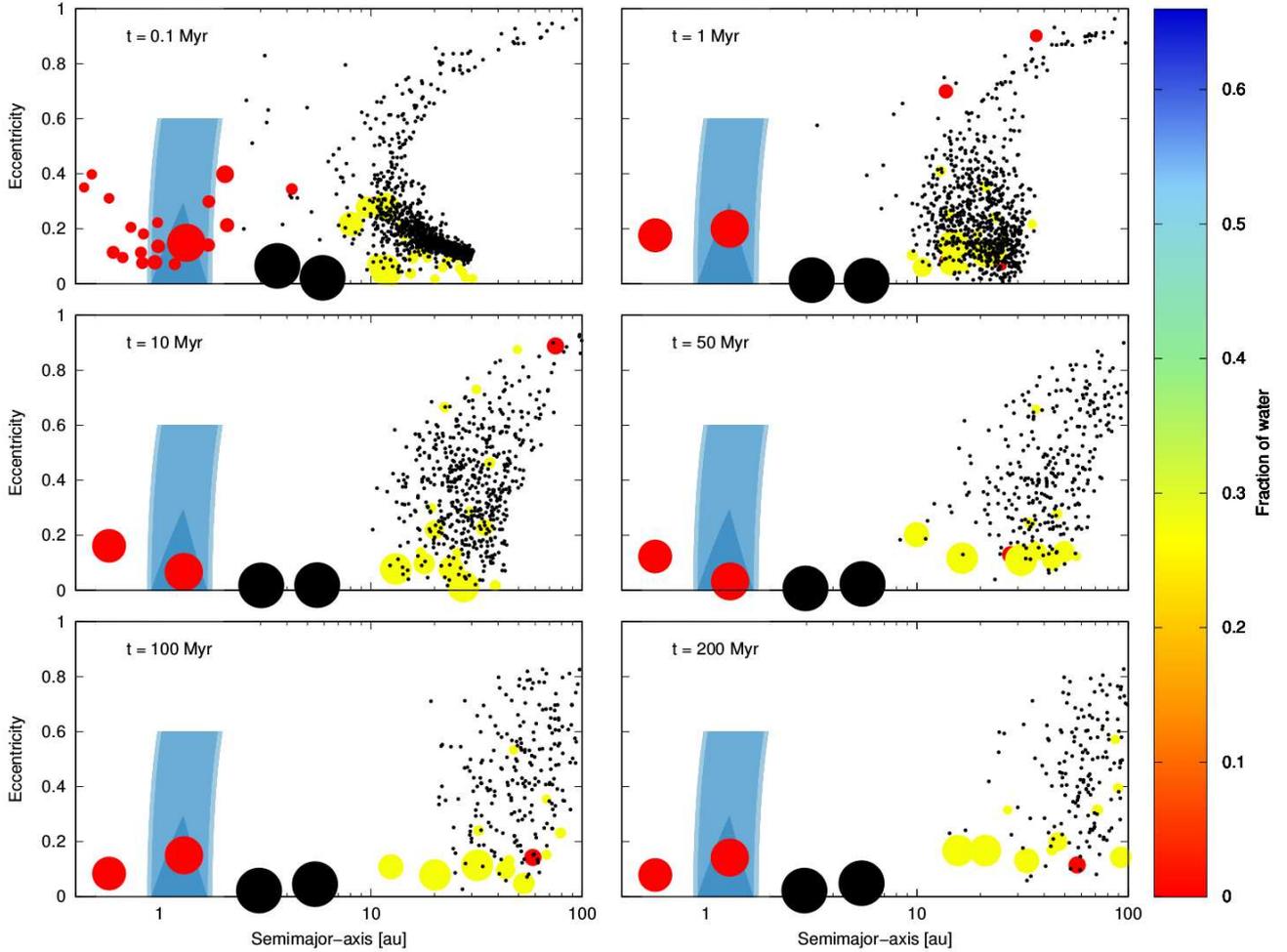}
  \caption{Time evolution of a simulation of the scenario with planetesimals of 10~km and without considering type I migration.
    The shaded areas represent the HZ regions already described in Section \ref{sec:sec1}. Planets are plotted as colored circles,
    planetesimals are plotted with black dots and the giant planets, which are Jupiter-like planets in this case, are represented in black.
The size of the Jupiter-like planets is fixed and is not in the same scale for the rest of the planets. The same happens with the
      planetesimal population, each planetesimal adopts a size of half the size of the smallest embryo, independently of its mass. Then, the size
      for the rest of the planets is represented by a lineal relation between the mass and the size, where the size for the smallest embryo is the double
      of the size of the planetesimals, and the size of the biggest planet is smaller that the size for a Saturn-like planet, like in the previous scenario.}
The color scale represents the fraction of water of the planets with respect to their total masses. This scenario form 
    dry PHPs within the HZ.
  \label{fig:fig5}
\end{figure*}

{\bf General dynamical evolution:} From the begining, the inner embryos are quickly excited by the two Jupiter-like gravitational perturbations and
by the perturbations with each other. The planetesimal population, which remains behind the gas giants, is also perturbed. However, very few planetesimals are able to
cross the giant barrier. Therefore, inner planetary embryos grow almost only by the accretion of other embryos within the first Myr of evolution. As an example of this
scenario figure \ref{fig:fig5} shows six snapshots in time of the semimajor-axis eccentricity plane of SIM$_1$. 

{\bf Mass remotion:} During the evolution of the systems a $\sim$ 8\% of the initial embryo population collides with the 
central star, a 19.8\% is accreted, and a $\sim$ 37.7\% is ejected from the system, considering the whole group of simulations. 
The planetesimals are mostly ejected from the systems ($\sim$ 77.1\%) due to the interactions with the giants and the outer planets. Only a few collide with the
central star (1.7\%) or are accreted (1.3\%) by the surrounding planets. In this scenario, as in the previous one, 
the planetesimal accretion is also a very inefficient process along the entire length of the disc.

{\bf Inner region:} Inside the orbit of the giant planets we find between 1 and 2 planets. Considering the 10 developed simulations the trend shows that is more
likely to find 2 planets instead of 1. These planets are completely dry Super-Earths or dry Mega-Earths with masses in a range of
$3.56\text{M}_\oplus$ to $15.96\text{M}_\oplus$. However, the trend is to find planets more massive than $10\text{M}_\oplus$.
In this scenario, 5 of 10 simulations formed one PHP. Although most of them are class A, all of them are completely dry because their accretion seeds come from the region inside the snowline. Their masses range between $3.56M_\oplus$ and $11.57M_\oplus$. The properties of these PHPs can be seen in table \ref{tab:tab5}.

{\bf Outer region:} Beyond the gas giants, planets and planetesimals survived until 400~au - 500~au. In these systems, the mix of material is less 
efficient than for planetesimals of 100~km. We only find one dry planet in the outer zone, beyond the giants, in half of the simulations.\\
$\bullet$ It seems that the gravitational interactions between the gas giants and the embryos in this case are stronger than with planetesimals of 100~km and
instead of injecting dry embryos into the outer zone, they directly eject them out of the system.

Finally, at the end of the evolution, the planetary systems resulting from S$_6$ in figure \ref{fig:fig1} are formed mostly by two dry Super or Mega-Earths, two Jupiter-like planets between $\sim$ 2.85~au and 5.78~au, and by several planets beyond these giants. In 5 of 10 simulations we find a dry PHP.

\begin{table}
  \centering
  \caption{Principal properties of the planets that remain within the HZ in scenarios with planetesimals of 10~km and without 
    type I migration. The row in bold corresponds to the results of Fig. \ref{fig:fig5}.}
  \label{tab:tab5}
  \begin{tabular}{l|c|c|c|c|c|} 
    \hline
    SIM & $a_{\text{i}}$ [au] & $a_{\text{f}}$ [au] & Mass $[\text{M}_\oplus]$ & W $[\%]$ & Habitable \\
        &                   &                 &                  &          & class \\
     \hline
    {\bf SIM$_1$} & {\bf 1.85}   & {\bf 1.30} & {\bf 10.80} & {\bf 0} & {\bf A} \\  
    SIM$_2$ & ---    & ---  & ---   & --- & --- \\
    SIM$_3$ & ---    & ---  & ---   & --- & --- \\
    SIM$_4$ & 1.22   & 1.01 & 14.41 & 0   & B \\
    SIM$_5$ & ---    & ---  & ---   & --- & --- \\
    SIM$_6$ & ---    & ---  & ---   & --- & --- \\
    SIM$_7$ & ---    & ---  & ---   & --- & --- \\
    SIM$_8$ & 1.99   & 1.17 & 6.86  & 0   & A \\
    SIM$_9$ & 2.15   & 1.45 & 11.57 & 0   & A \\
    SIM$_{10}$ & 0.57 & 0.94 & 3.56  & 0   & C \\
    \hline
  \end{tabular}
\end{table}

\subsubsection{Scenarios with planetesimals of 1~km}

{\bf Main perturbers:} A  Saturn-like planet and a Jupiter-like planet act as the main perturbers of these systems. The Saturn-like planet is 
the gas giant which is closer to the central star and is initially located at 4~au. This planet finishes located between 1.68~au and 2.97~au due to an embryo-driven
migration after 200 Myr of
evolution and presents a mean mass of $46\text{M}_\oplus$ ($0.48\text{M}_{\text{S}}$). This inward migration is significant but it is not
enough to disperse all the solid material of the inner zone of the disc as it does, for example, the Saturn-like planet of the scenario
of 100km. The Jupiter-like planet, originally located at 5.67~au, finishes always located
around $\sim$ 4.6~au with a mass of $1.6\text{M}_{\text{J}}$.\\
$\bullet$ {\bf Icy giants:} Unlike the previous scenarios, this one presents two Neptune-like planets\footnote{From PI, the adopted definition for an icy giant planet is that of a planet which presents a mass 
of the envelope greater than 1\% of the total mass, but lower than the mass of the core. Particularly, if the mass of the envelope is 
greater than 1\% of the total mass and lower than 30\% of the total mass, we consider it is a Neptune-like planet.} between the
rocky embryos and the gas giant planets, at the beginning of the simulations (see S$_{4}$ in Fig. \ref{fig:fig1}) with masses of $\sim$ $15M_\oplus$ and $\sim$ $17M_\oplus$.

\begin{figure*}
  \includegraphics[angle=270, width=\textwidth]{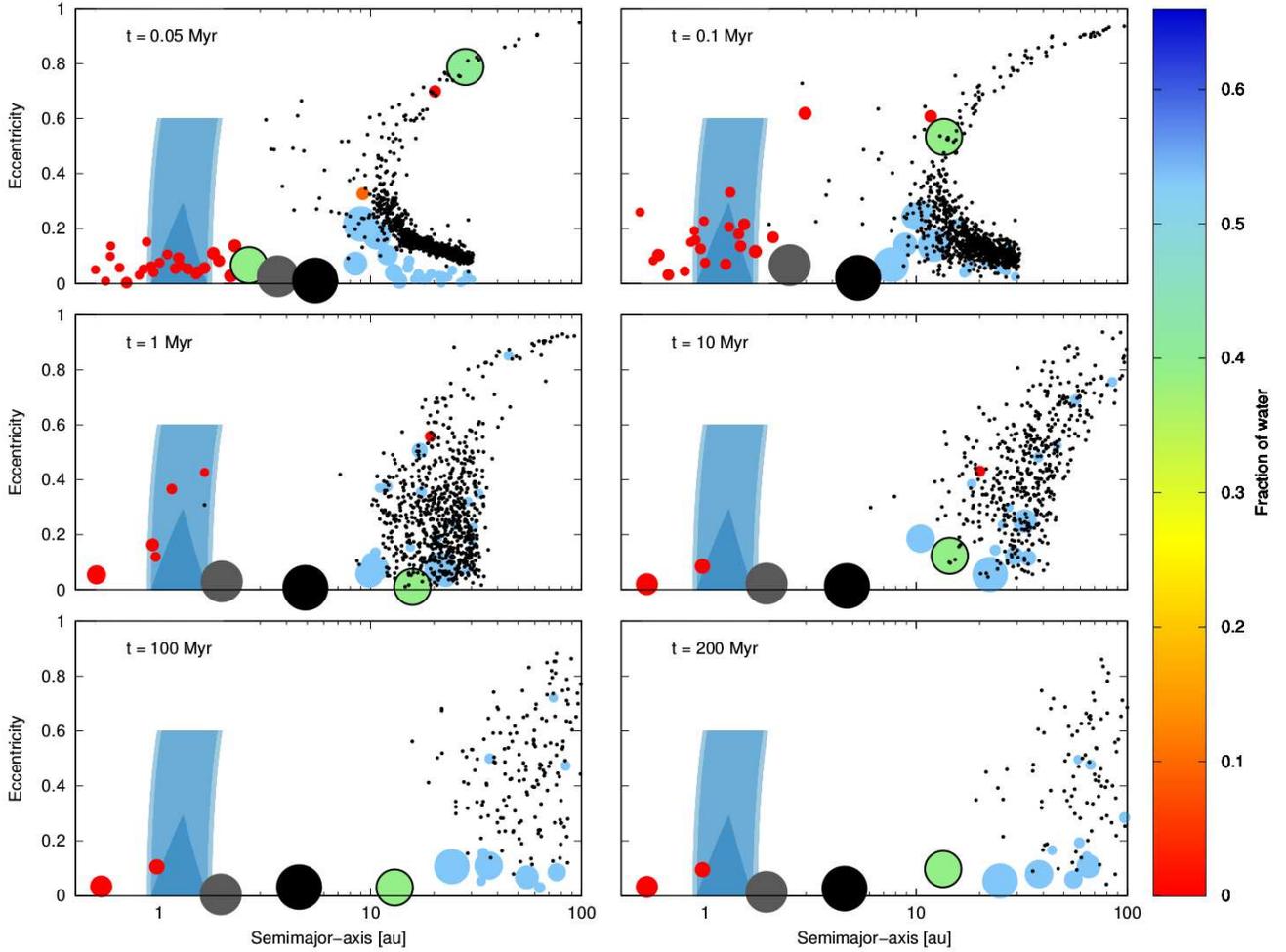}
  \caption{Time evolution of SIM$_{7}$ of the scenario with planetesimals of 1~km and without considering type I migration.
    The shaded areas represent the HZ regions already described in Section \ref{sec:sec1}. Planets are plotted as colored circles,
    planetesimals are plotted with black dots and the giant planets, which are a Saturn-like planet and a Jupiter-like planet in 
    this case, are represented in grey and black, respectively. The two big colored planets with black borders represent the
    two Neptune-like planets. The sizes of the Saturn-like and Jupiter-like planets are fixed and are not in the same scale
      for the rest of the planets. The same happens with the planetesimal population, each planetesimal adopts a size of half the size of the smallest embryo, independently of its mass.
      Then, the size for the rest of the planets is represented by a lineal relation between the mass and the size, where the size for the smallest embryo is the double
      of the size of the planetesimals, and the size of the biggest planet is smaller that the size for a Saturn-like planet, like in the previous scenario.
    The color scale represents the fraction of water of the planets with respect to their total masses. 
    This scenario form 4 PHPs within the HZ, but 3 of those 4 are the Neptune-like planets. In this particular case the planet that remain within
    the HZ is a Super-Earth of $2.55\text{M}_\oplus$ classified as a PHP class B. 
  }
  \label{fig:fig6}
\end{figure*}

{\bf General dynamical evolution:} From the beginning, the embryos are excited due to the mutual perturbations between them 
and with both, the icy and the gas giants. The first most common result, found in 6 of 10 simulations, is that one of the two Neptunes is ejected from the systems
within the first $\sim$ 0.1~Myr. The second most common result, which happens in 5 of the 6 simulations mentioned before, is that the other Neptune
is scattered towards the outer zone of the disc presenting encounters with an excited planetesimal population. Here it can be seen the importance of the dynamical
friction produced by the planetesimals that dampens the eccentricity and inclination of the Neptune-like planet. The mass available in planetesimals to produce
this effect is, during this process, high enough to cause the planet to settle in the mid-plane of the disc in less than $\sim$ 1 Myr. We can see in figure \ref{fig:fig6} six snapshots in time of the semimajor-axis eccentricity plane of SIM$_{10}$ as an example of this scenario.
Although is not the most common result, we find that in 3 of the 10 simulations one of the Neptune-like planets survives inside the giants locations
and finishes within the HZ while the other one is ejected. 

{\bf Mass remotion:} During the simulations, a 15.8\% of the initial embryo population collides with the central star, a 12.7\% 
is accreted, and a 46.73\% is ejected from the system. While only between 1 and 2 planets remain inside the giants, the rest of them are located 
beyond them. The planetesimals, perturbed by the giants and outermost planets, are mainly ejected from the system ($\sim$ 84.4\%). 
Only a few collide with the central star (0.6\%) or are accreted (0.8\%) by the surrounding planets. Only a $\sim$ 14.0\% of the 
initial planetesimals survives at the end of the simulations with semimajor-axis almost greater than 20~au and mean high eccentricities.
Here again, the planetesimal accretion is an inefficient process, above all, in the inner region of the disc, since no planet originally 
located in the inner zone has succeeded in accreting planetesimals coming from beyond the giants.

{\bf Inner region:} The inner zone of the disc in general presents one or two planets. The trend shows that is more likely to find one, which in almost all the cases is a Super-Earth. However, in 3 of the 10 simulations one of inner planets is a Neptune-like planet. With the exception of the Neptune-like planets, the inner planets are completely dry and present masses in a range between $1.1\text{M}_\oplus$ and $6.18\text{M}_\oplus$. In general, this scenario achieved to form one PHP in 4 of 10 simulations but, as we mentioned before, 3 of that 4 PHPs are Neptune-like planets, which present high water contents. The other PHP is a dry Super-Earth of $2.55\text{M}_\oplus$. The properties of these PHPs can be seen in table \ref{tab:tab6}.

{\bf Outer region:} Beyond the giants, there is a planet and a planetesimal population until $\sim$ 400~au. In most of the simulations the planets in this
outer region of the disc are water-rich planets, and we do not see planets from the inner zone. Moreover, in most of the cases, one of these
planets is one of the Neptune-like planets, originally located between 2.6~au and 2.9~au.

\begin{table}
  \centering
  \caption{Principal properties of the planets that remain within the HZ in scenarios with planetesimals of 1~km and without type I 
migration. The planets with $^*$ are the Neptune-like planets. The row in bold corresponds to the results of Fig. \ref{fig:fig6}.}
  \label{tab:tab6}
  \begin{tabular}{l|c|c|c|c|c|} 
    \hline
    SIM & $a_{\text{i}}$ [au] & $a_{\text{f}}$ [au] & Mass $[\text{M}_\oplus]$ & W $[\%]$ & Habitable \\
        &                   &                 &                  &          & class \\
     \hline
    SIM$_1$ & ---   & ---  & ---   & --- & --- \\  
    SIM$_2$ &  ---  & ---  & ---   & --- & ---\\
    SIM$_3$ & 2.92  & 1.39 & $15.37^*$ & 39.10 & A \\
    SIM$_4$ & 3.43  & 1.14 & $17.32^*$ & 38.91 & A \\
    SIM$_5$ & ---   & ---  & ---   & --- & --- \\
    SIM$_6$ & ---   & ---  & ---   & --- & --- \\
    {\bf SIM$_7$} & {\bf 1.07}  & {\bf 0.97} & {\bf 2.55}  & {\bf 0}  & {\bf B} \\
    SIM$_8$ &  ---  & ---  & ---   & --- & --- \\
    SIM$_9$ & ---   & ---  & ---   & --- & --- \\
    SIM$_{10}$ & 3.43 & 1.41 & $17.32^*$ & 38.91 & A \\
    \hline
  \end{tabular}
\end{table}

Finally, at the end of the simulations, the planetary systems resulting from S$_4$ in figure \ref{fig:fig1} are formed by one or two inner planets which are dry Super-Earths 
or water-rich Neptune-like planets. A Saturn-like is followed by a Jupiter-like planet, and other several planets and planetesimals are beyond them. 
We find PHPs in 4 of 10 simulations. One of them is a dry Super-Earth and the other 3 are
Neptune-like planets that were not scattered. The common result regarding these
Neptunes in the HZ, is that they did not suffer collisions either from embryos or from planetesimals, reason why they could have 
conserved their primordial envelopes during all the evolution. Thus, although their final locations 
define them as PHPs, it is, in fact, very likely that they are not. Therefore, this scenario is not really interesting. 

\subsubsection{Scenarios with planetesimals of 100~m}

{\bf Main perturbers:} These simulations present two Jupiter-like planets of $1.49\text{M}_{\text{J}}$ and $5\text{M}_{\text{J}}$,
located at around $\sim$ 9.9~au and $\sim$ 19.3~au during all the evolution, respectively, as the main perturbers of the system.\\
$\bullet$ {\bf Icy giants:} Two Neptune-like planets of $\sim$ $13\text{M}_\oplus$ and $\sim$ $15\text{M}_\oplus$ are located between the 
inner embryo distribution and the Jupiter analogues, at 6.82~au and at 7.80~au, respectively, at the beginning of the simulations 
(see S$_{1}$ in Fig. \ref{fig:fig1}).

{\bf General dynamical evolution:} As from the begining, the inner embryos get excited very quickly acquiring very high 
eccentricities. This causes that the orbits of these objects cross each other, leading to accretion collisions, ejections and
collisions with the central star.\\
$\bullet$ From the beginning, these collisions may occur between dry and water-rich embryos since these 
two different populations are not separated by a big perturber as in the previous scenarios.

\begin{figure*}
  \includegraphics[angle=270, width=\textwidth]{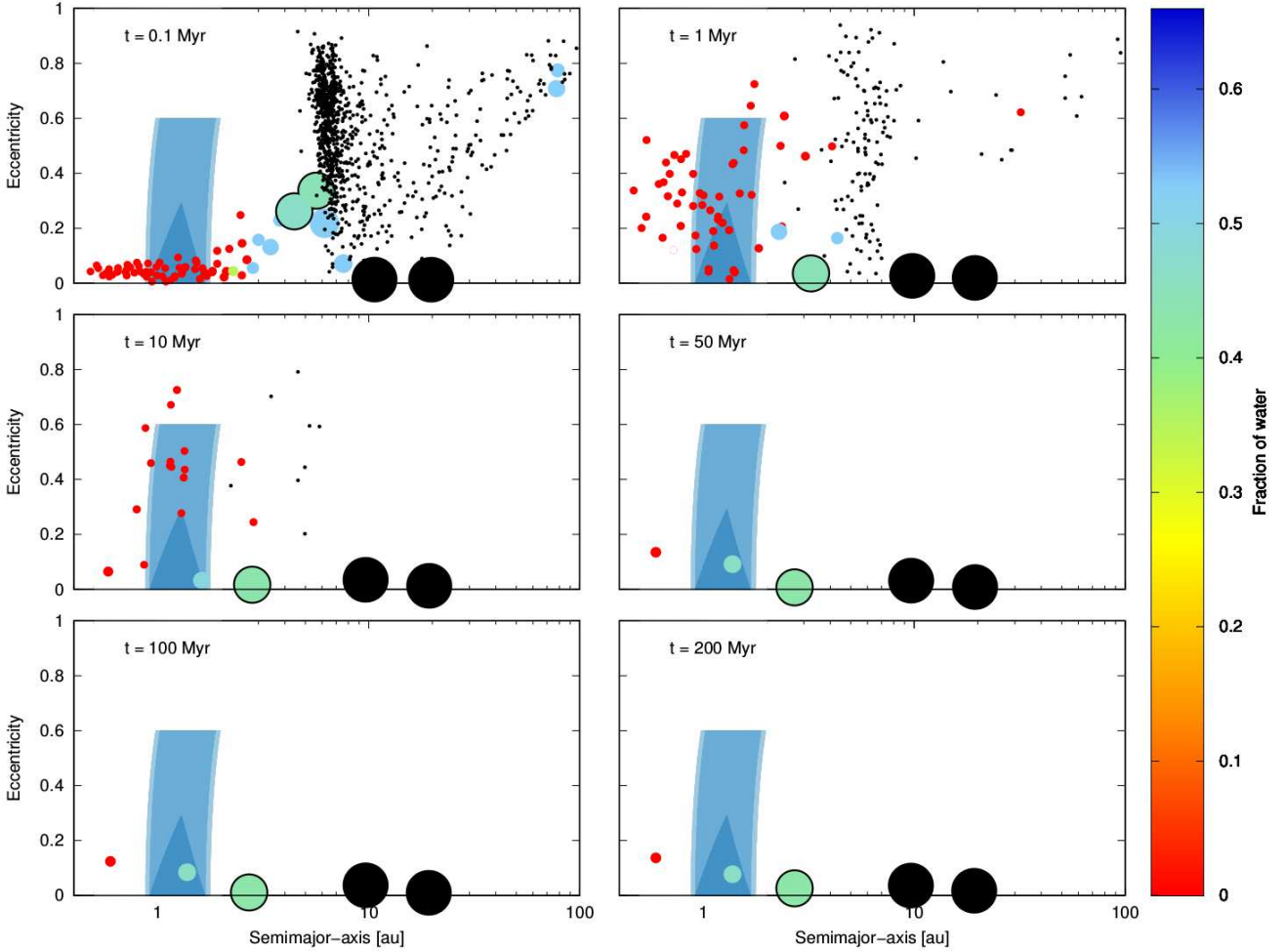}
  \caption{Time evolution of SIM$_{10}$ of the scenario with planetesimals of 100~m and without considering type I migration.
    The shaded areas represent the HZ regions already described in Section \ref{sec:sec1}. Planets are plotted as colored circles,
    planetesimals are plotted with black dots, the two big colored circles with black borders are Neptune-like planets and the two
    black circles represent Jupiter analogues of $1.49\text{M}_{\text{J}}$ and $5\text{M}_{\text{J}}$. The size of the Jupiter-like planets is fixed and is not in the same scale
      for the rest of the planets. The same happens with the planetesimal population, each planetesimal adopts a size of half the size of the smallest embryo, independently of its mass.
      Then, the size for the rest of the planets is represented by a lineal relation between the mass and the size, where the size for the smallest embryo is the double
      of the size of the planetesimals, and the size of the biggest planet is smaller that the size for a Saturn-like planet, like in the previous scenario.
    The color scale represents the fraction of water of the planets with respect to their total masses. This is the first scenario that form water-rich PHPs.}
  \label{fig:fig7}
\end{figure*}

The fate of the Neptune-like planets is always the same. The trend shows that one of them is ejected from the system
within the first million year of evolution, while the other one, which in most of the cases the inner Neptune, stays in the inner zone. The most common result 
among the simulations is that this Neptune-like planet do not suffer collisions of any kind. Thus, it could retain its primitive atmosphere.
We show in Fig. \ref{fig:fig7} six snapshots in time of the evolution of SIM$_{10}$, as an example of this scenario. The initial conditions
correspond to S$_{1}$ in Fig. \ref{fig:fig1}.
Although in these scenarios the planetesimal population resides in the inner zone of the disc, they are initially located between
5~au and 7~au, just beyond one of the Neptunes. Thus, this population does not directly interact with the inner embryo population 
between 0.5~au and 4~au. 

{\bf Inner region:} The inner region of the disc presents between 2 and 3 planets and in all the simulations one of these inner planet is one of the 
Neptune-like planets. The others are dry or water-rich Earths, Super-Earths or Mega-Earths with masses in a range between $1.02\text{M}_\oplus$ and $9.10_\oplus$.\\
$\bullet$ This scenario forms 6 PHPs, 5 of them are, unlike the ones in previous scenarios, Super-Earths with very high water contents what indicates that their accretion seeds come from beyond the snowline.\\
The masses of these PHPs are between $2.64M_\oplus$ and $7.14M_\oplus$. The most interesting characteristics of
these PHPs can be seen in table \ref{tab:tab7}.

{\bf Outer region:} This scenario does not present any kind of objects, nor planets nor planetesimals, beyond the giants
after 200 Myr of evolution. \\
$\bullet$ Planetary systems in this case extend only up to the position of the outermost Jupiter
analogue at $\sim$ 20~au. This is a direct consequence
of the initial conditions which do not present an embryo and planetesimal population beyond the gas giants. Besides, the strong gravitational 
interactions between the two gas giants with the inner embryo and planetesimal population only manage to eject bodies outside the
systems and not to disperse them towards the outer zones of the disc.

{\bf Mass remotion:} In this scenario, and considering all the simulations, a $\sim$ 50\% of the embryos collides with the central star, a $\sim$ 35\%
is ejected from the system and a $\sim$ 23\% is accreted by other embryos.  The planetesimal population is 
excited mostly by the gas giants but also by the Neptune-like planet, and in only $\sim$ 20 Myr this population is completely removed from 
the system. \\
$\bullet$ As, unlike the other scenarios the planetesimals are initially located inside the giants, an important percentage of them collide with
the central star ($\sim$ 11\%) while the great majority is ejected from the system (89\%). \\
Only a 0.3\% of the planetesimals are accreted during the whole
evolution of the systems. Thus,
although the location of these bodies is different from the other scenarios, the accretion of planetesimals is still a very inefficient process.

\begin{table}
  \centering
  \caption{Principal properties of the planets that remain within the HZ in scenarios with planetesimals of 100~m and without 
    type I migration. The row in bold corresponds to the results of Fig. \ref{fig:fig7}.}
  \label{tab:tab7}
  \begin{tabular}{l|c|c|c|c|c|} 
    \hline
    SIM & $a_{\text{i}}$ [au] & $a_{\text{f}}$ [au] & Mass $[\text{M}_\oplus]$ & W $[\%]$ & Habitable \\
        &                   &                 &                  &          & class \\
     \hline
    SIM$_1$ & 4.91 & 1.26 & 3.69 & 46.36 & A \\  
    SIM$_2$ & 6.82 & 1.71 & 7.14 & 52.28 & B \\
    SIM$_3$ &  ---  & ---  & ---   & --- & ---\\
    SIM$_4$ & 5.28  & 1.54 & 3.63 & 52.70 & B \\
    SIM$_5$ & ---   & ---  & ---   & --- & --- \\
    SIM$_6$ & ---   & ---  & ---   & --- & --- \\
    SIM$_7$ &  ---  & ---  & ---   & --- & --- \\
    SIM$_8$ & 4.26  & 1.62 & 4.39  & 51.16 & B \\
    SIM$_9$ & 4.26  & 1.66 & 2.64  & 52.90 & B \\
    {\bf SIM$_{10}$} & {\bf 4.57} & {\bf 1.38} & {\bf 3.34} & {\bf 46.28} & {\bf A} \\
    \hline
  \end{tabular}
\end{table}

In conclusion, the planetary systems of this scenario resulting from S$_1$ in figure \ref{fig:fig1} present between 2 and 3 planets in the inner zone of the disc, one of which is a 
Neptune-like planet. In 6 of 10 simulations we find a water-rich Super-Earth PHP class A or B. Then we find a Neptune-like planet around 3~au and 2 Jupiter analogues 
beyond $\sim$ 10~au. This scenario results to be interesting since formed PHPs 
with high primordial water contents.

\subsection{Comparison between planetesimal sizes}

\begin{figure}
  \includegraphics[angle=270, width=\columnwidth]{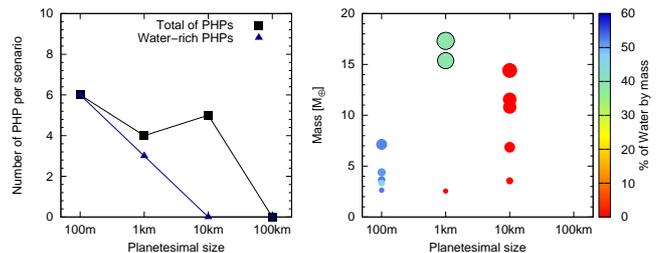}
  \caption{Total number of formed PHPs and number of water-rich PHPs per scenario as a function of the planetesimal size (left panel) and masses of the formed PHPs per planetesimal size.}
\label{fig:fig8}
\end{figure}

Here we highlight the most important differences and similarities between the reference scenarios formed by planetesimals of different sizes and
without migration rates during the gaseous phase, focusing mainly on the inner regions where potentially habitable planets form.
\begin{itemize}
\item Following panels S$_{1}$, S$_{4}$, S$_{6}$ and S$_{9}$ in figure \ref{fig:fig1} it can be seen that the mass of the main perturbers of the different
  scenarios grows as the size of the planetesimals decreases. During the evolution of the N-body simulations, those less gas giant planets, like the Saturn-like planet
  in scenarios of 100~km and the Saturn-like planet in scenarios of 1~km, suffered a significant inward embryo-driven migration. This migration particularly
  placed the Saturn-like planet in scenarios of 100~km at the outer edge of the HZ, preventing the formation of PHPs. 
\item Regarding the global configuration of the different scenarios at the end of the gaseous phase we can see that scenarios formed with planetesimals of 100~m are
quite different from those formed with planetesimals of 1~km, 10~km and 100~km. 
Just by looking at the initial conditions in Fig. \ref{fig:fig1}, and comparing S$_{1}$ with S$_{4}$, S$_{6}$ and S$_{9}$, it can be seen that: 1) the planetesimal 
population is located inside the giants positions, instead of outside, 2) the gas giants are much further away from the central 
star than they are in the other scenarios, and 3) a water-rich embryo population is directly next to the dry one, in the inner region
of the disc. These differences in the initial conditions conduct to different results in the final configurations of the N-body
simulations.
\item The mass remotion mechanisms, this is embryo and planetesimal remotion via accretion, collisions with the central star and via ejections of the systems, are
  quite similar in all the scenarios being the most efficient mechanism the ejection, both of embryos and planetesimals. The remotion of planetesimals is completely
  efficient for the scenario formed by planetesimals of 100~m.
\item Unless the outer regions of the scenario formed by planetesimals of 100~m, which do not present planets or planetesimals at all, the other three scenarios present
  a mixed population of water-rich and some dry planets, and planetesimals untill 200~Myr of evolution.  
\item The inner regions of all the scenarios are generally configured quickly, within the first 1 Myr of evolution for scenarios of 100~km and 10~km, and within the
  first 10 Myr and 50 Myr for scenarios of 1~km and 100~m, respectively.
\item The efficiency in the formation of PHPs seems to be higher for scenarios formed from small planetesimals than from bigger ones,
  being completely null for planetesimals of 100~km. The final masses of these PHPs are in a wide range of values. The masses of the PHPs of the scenario formed from planetesimals of 100~m are in the
  range of the Super-Earths, while the masses of the PHPs in the other scenarios are in the range of the Super and Mega-Earths. It is important to remark that,
  in scenarios with planetesimals of 1~km, 3 of the 4 PHPs founded where Neptune-like planets that do not present any interest. Regarding water contents, the
  only PHPs of real interest that present water contents are those formed in scenarios with planetesimals of 100~m. Figure \ref{fig:fig8} shows these mentioned
  characteristics of the PHPs formed within all the four reference scenarios. The left panel shows the total number of formed PHPs per scenario as a function of the planetesimal size
  and the right panel shows their masses per planetesimal size.
\end{itemize}

\subsection{Results sensitivity to scenarios with type I migration}
\label{subsec:subsec6}

The previous sections described the general results of the default N-body simulations performed considering different planetesimal
sizes and no type I migration. Since one of the important results of PI was that the most favorable scenarios for
the formation of SSAs were planetary systems with low/null type I migration rates, we also performed N-body simulations for those
SSAs at the end of the gaseous phase, that presented type I migration rates reduced to a 1\% and to a 10\%. The goal of this 
section is not to describe in detail, as we did before, the particular results of each of these simulations, but to give a global view of 
what are the effects that this phenomenon produces and how the previous results could change because of it.

We performed 10 different simulations for each scenario S$_{2}$, S$_{3}$, S$_{5}$, S$_{7}$, S$_{8}$ and S$_{10}$, that represent planetary
systems with type I migration rates reduced to a 1\% and to a 10\%. Those simulations with type I migration reduced to a 1\% do not present
significant differences from the ones without type I migration. For example, the initial conditions of S$^*_9$ (100~km without type I migration)
and S$^*_{10}$ (100~km with type I migration reduced to 1\%) are very
similar and thus, the final configurations of the N-body simulations are similar too. Although the mass of the Saturn-like planet of
S$^*_{10}$ is smaller than the Saturn-like planet of S$^*_9$ in a factor $\sim$ 3, the main result is that none of these scenarios form planets in the 
inner region of the disc. Something similar occurs between S$_6$ (10~km without type I migration) and S$_{7}$ (10~km with type I migration reduced to 1\%) although S$_6$
presents 2 gas giants as main perturbers and S$_{7}$ presents only 1. The main results show
that the inner region of the disc presents between 1 and 2 dry planets with masses in the range of the Super and Mega-Earths, and that
some of them, between 3 and 5, are completely dry PHPs. In the same way, the results concerning scenarios of 1~km with type I
migration reduced to a 1\% and without type I migration, are also similar: the main result shows that both scenarios form dry PHPs.
Regarding scenarios of 100~m, both kinds of simulations, with migration reduced to a 1\% and without type I migration, form water-rich PHPs.

The results of PI showed that only very few planetary systems formed from planetesimals of 100~m and 10~km were able to form SSAs at the
end of the gaseous phase with type I migration rates reduced to a 10\%. Although this is an uncommon result, we still performed 
simulations for this scenarios, whose initial conditions are represented by S$_3$ (for scenarios with planetesimals of 100~m) and by 
S$_8$ (for scenarios with planetesimals of 10~km).

For scenarios with planetesimals of 10~km, the type I migration during the gaseous-phase, favors the giant planet to form closer 
to the central star. In this case, this result is detrimental to the formation of PHPs. However, for scenarios with planetesimals of 
100~m it seems to be just the other way around. We will describe these situations in more detail below.

\begin{figure*}
  \includegraphics[angle=0, width=\textwidth]{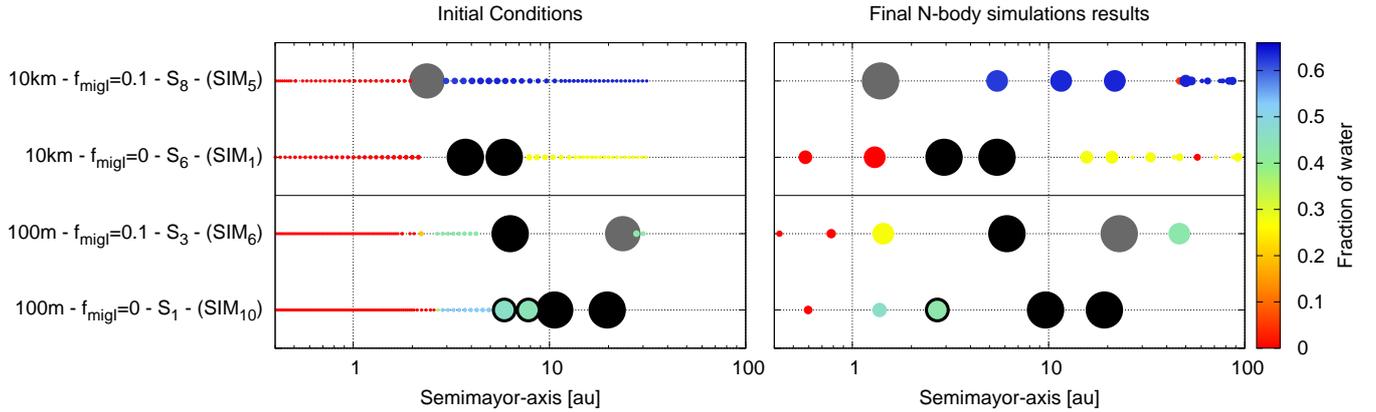}
  \caption{Initial conditions (left panel) and final N-body simulations results (right panel) of scenarios formed with planetesimals of 
    100~m and 10~km, without type I migration ($f_{\text{migI}}=0$) and with a type I migration reduced to a 10\% ($f_{\text{migI}}=0.1$).}
\label{fig:fig9}
\end{figure*}

Figure \ref{fig:fig9} shows initial and final configurations of planetary systems formed from planetesimals of 100~m (bottom) 
and 10~km (top). The left panel shows the initial conditions at the beginning of the N-body simulations and the right panel shows 
the N-body simulations results. On the one hand, comparing the two scenarios of 10~km, with and without type I migration, we can see 
that type I migration during the gaseous stage, leads the giant planet of the system to locate closer to the central star 
than the inner giant is in the scenario without migration. This result, which is a natural consequence of type I migration, and the 
fact that the position of the giant could also
change during the post-gas evolution of the system due to an induce migration via the gravitational interactions with the surrounding 
embryos and planetesimals, is unfavorable for the formation of PHPs. In fact, we already described that the simulations with 10~km 
planetesimals and no type I migration were able to form 5 dry PHPs. However, the 10 simulations developed with 10~km planetesimals and 
type I migration rates reduced to a 10\% were not able to form PHPs at all.
On the other hand, comparing the two scenarios of 100~m, with and without type I migration, we can see that in spite of the fact that 
type I migration manages to locate the giant further inside, regarding the position of the innermost giant in the scenario without 
type I migration, in this case, this effect favors the formation of PHPs. In fact, as we have already comment in section 
\ref{subsec:subsec5}, the simulations with planetesimals of 100~m and no type I migration were able to form 6 water-rich PHPs. However, 
the 10 simulations developed with type I migration rates reduced to a 10\% were able to form 11 PHPs, 9 are classified as
PHPs class A and 2 class B. The final masses of these PHPs are between $0.93\text{M}_\oplus$ and $3.73\text{M}_\oplus$, with water contents between
24.3\% to 33.5\%. It would seem then, that there is a preferential location for the inner giants, in terms of the efficiency in the 
formation of PHPs. We will return to this point later. 

To conclude, scenarios with type I migration rates highly reduced, do not present significant differences from those scenarios without type
I migration. However, scenarios with type I migration rates that represent a 10\% of the \citet{Tanaka2002} type I migration rates, 
show different results for planetary systems formed from small and big planetesimals. This phenomena seems to favor the formation
of PHPs in scenarios with small planetesimals, but definitely avoids their formation in scenarios formed from big planetesimals.

\subsection{Results sensitivity to scenarios with gap-opening giants}
\label{subsec:subsec7}

Another important phenomenon that could modify the final configuration of the planetary systems, is the type II 
migration. The type II migration occurs when a planet is massive enough to fulfill the \citet{Crida2006} criteria, therefore to open a gap in the disc, which depends not only on the mass and the position of the planet but also on the Hill radius of the planet,
  the viscosity and the height scale of the disc (see PI for more details). This phenomenon, which was included in the model of planet formation of PI, produced an inward migration
  during the gaseous phase, only for this gap opening giant, and an increase in its mass due to the accretion of gas and other planets in its path. It is important to remark that the location of the gas giant at the end of the gaseous phase is not a consequence due only to
the type II migration, but because once type II migration is turned on, the planet, which moves inward, accretes other planets in its path.
Recall from PI that the resulting semimajor-axis of a fusion between two planets is given by the conservation of angular momentum (see eq. 33 in PI).

It has been proved by many models that an inward giant planet migration shepherds some planets captured at inner mean motion resonances with
  the giant, which are then dispersed into outer orbits, leading to the formation of planetary systems with an inner compact planet
  population and an outer much more scattered planet population \citep{FoggNelson2005,FoggNelson2006,Raymond2006,FoggNelson2007a,Mandell2007}.
  Particularly, \citet{FoggNelson2009} have showed that including a viscous gas disc with
  photoevaporation in an N-body code and although the type II inward migration, it is still possible to form planets within the HZ in low-eccentricity warm-Jupiter systems if the giant planet makes a
  limited incursion into the outer regions of the HZ.
  Our model of planet formations does not yet include the scattering effects from a migrating giant planet or the gravitational interactions
  between planets that can lead to resonance captures. Therefore, the results of our N-body simulations could be affected by the initial conditions which could had overestimated the number of
  planets in the inner zone of the systems at the end of the gaseous phase.

The main difference between the opening gap giant planets with those which did not, is their final mass. Gap opening planets are, 
in general, more massive than those that did not. Table \ref{tab:tab8} shows the ranges 
of masses in scenarios with different planetesimal sizes for those giants that opened and not opened a gap. It is important to note 
that this table was drawn up from the initial conditions of Fig. \ref{fig:fig1} and from 4 more initial conditions for the different 
sizes, then used to run more N-body simulations, which present at least one gas giant that managed to open a gap in the disc.

\begin{table}
  \centering
  \caption{This table shows the range of masses for the giant planets that opened and did not opened a gap, 
in the scenarios with planetesimals of different size. The ranges are in Saturn and Jupiter masses due to the fact that we find both 
types of planets in scenarios with 100~m, 1~km and 10~km. As it can be seen in the table, the masses of those giants that opened a gap
in the disc during the gaseous phase is greater than the other ones. }
  \label{tab:tab8}
  \begin{tabular}{l|c|c|c|} 
    \hline
    Scenario & Giants that    & Giants that \\
             & did not opened a gap      & opened a gap \\
     \hline
    100~m    & $0.81\text{M}_{\text{S}} - 5\text{M}_{\text{J}}$ & $5.63\text{M}_{\text{J}} - 8.92\text{M}_{\text{J}}$ \\
    1~km     & $0.48\text{M}_{\text{S}} - 4\text{M}_{\text{J}}$ & $7.23\text{M}_{\text{J}}$ \\
    10~km    & $1.33\text{M}_{\text{S}} - 0.8\text{M}_{\text{J}}$ & $1.88\text{M}_{\text{J}} - 2.10\text{M}_{\text{J}}$ \\
    100~km   & $0.56\text{M}_{\text{S}} - 1.90\text{M}_{\text{S}}$ & $3.23\text{M}_{\text{J}}$ \\
    \hline
  \end{tabular}
\end{table}

\begin{figure*}
  \includegraphics[angle=0, width=\textwidth]{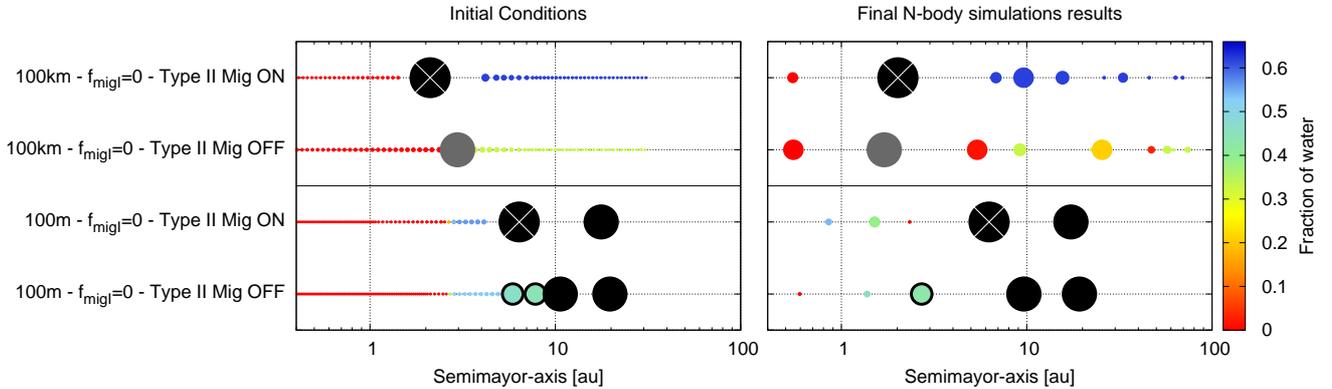}
  \caption{Initial conditions (left panel) and final N-body simulations results (right panel) of scenarios formed with planetesimals of
    100~m and 100~km, without type I migration (fmigI = 0) and with and without type II migration. The black points represent Jupiter-like planets
    while the grey points represent Saturn-like planets. The black points with a white cross are those giant planets that had managed to open a gap in the
  disc during the gaseous phase evolution and had migrated inward.}
\label{fig:fig10}
\end{figure*}

In order to see if there were significant differences between the default scenarios and scenarios without type I migration but with at least
one giant that managed to \emph{turn on} type II migration, we run more simulations. We chose one initial condition for each planetesimal 
size that presents a gas giant that has opened a gap. This is, we chose 4 new initial conditions and we run 10 simulations per each of these 4 
scenarios changing, as we did before, the seed. Figure \ref{fig:fig10} shows, in a similar way as figure \ref{fig:fig10} does, a comparison between simulations of the two
extreme reference scenarios of 100~km and 100~m without type I and type II migration and scenarios of 100~km and
100~m without type I migration but with the type II migration activated. The left panels show the initial conditions at the begining of the N-body simulations, this is, the final
results at the end of the gaseous phase, and the right panels show the N-body simulations results after 200 Myr of evolution.

A global result, found in all the new simulations, is that, due to the high mass of the new opening gas giants, those scenarios which present a
planet and planetesimal population beyond the gas giants, do not present a mix of dry and water-rich material in the outer zone of the disc.
This is, the inner dry embryos are directly ejected from the systems and are not scattered towards the outer regions.

Regarding the efficiency in forming PHPs, in scenarios with planetesimals of 100~km, it still remains completely null. None
of the simulations with planetesimals of 100~km, without type I migration and with a planet that opened a gap, managed to form PHPs. The gap opening Jupiter-like planet in the scenario
of 100~km of figure \ref{fig:fig10} (the big black point with a white cross) have $3.22M_{\text{J}}$ and is located around $\sim$ 2~au, while the Saturn-like planet in the scenario
of 100~km which did not opened a gap presents a mass of $0.5M_{\text{J}}$ and is located, initially at 2.96~au and finally at $\sim$ 1.7~au.   
For scenarios of 10~km, the production of PHPs in the 10 simulations performed with a planet that opened a gap is lower than in the default
scenarios. This scenario formed 2 dry PHPs instead of 4 with masses of $\sim$ $2M_\oplus$ and $11M_\oplus$. Following with scenarios of 1~km, in this case the PHP production was similar.
Both scenarios, with and without an opening gas giant, produced 5 PHPs with masses ranging from $0.9M_\oplus$ to $\sim$ $6M_\oplus$. Finally, scenarios with planetesimals of 100~m and a
giant planet that opened a gap formed 9 PHPs with masses between $5.28M_\oplus$ and $11M_\oplus$, 3 more than in the default case. In these scenarios, the gap opening Jupiter-like planet
  has a mass of $5.63M_{\text{J}}$ against the $1.49M_{\text{J}}$ of the Jupiter-like planet that did not opened a gap, and the first one is located around 6.38~au while the second one is
located at 10~au. This analysis of the comparison between the whole group of simulations, with and without a gap opening planet, 
shows us that the type II inward migration coupled with the increase of the mass of the giant seems to act in a similar way as type I migration do.
Although this is the outcome of our simulations, more simulations should be carried out to confirm this trend.
These effects seem to favor the formation of PHPs for scenarios with small planetesimals and to be detrimental for scenarios formed from
big planetesimals. 

\subsection{Possible range of mass and location of the inner giant to form PHPs}
\label{subsec:subsec7b}

\begin{figure}
  \includegraphics[angle=0, width=\columnwidth]{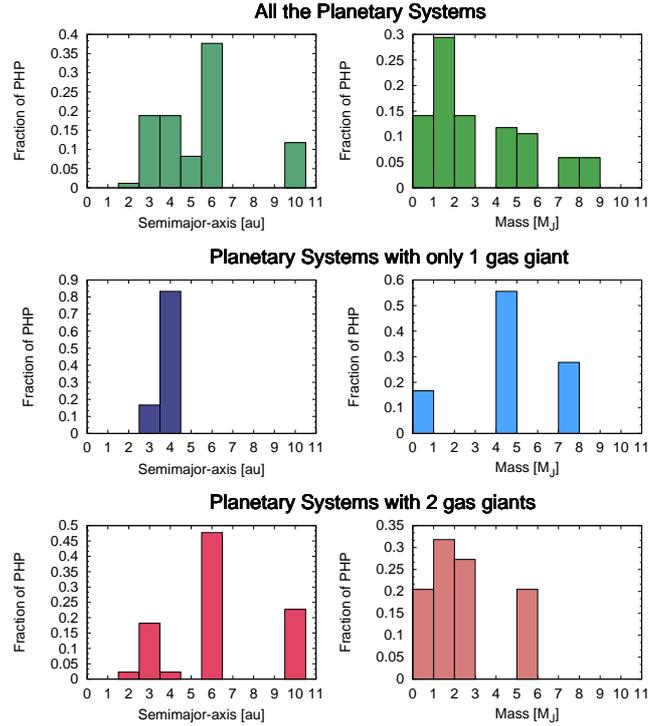}
  \caption{Histograms showing the fraction of PHPs formed, for all the simulations performed, as a function of the semimajor-axis of the 
innermost giant planet (top left panel) and as a function of the mass of the innermost giant planet (top right panel). 
The middle and the bottom panels show the same histograms but considering those performed simulations with only one gas giant and with 
two giants, respectively.}
\label{fig:fig11}
\end{figure}

With the aim of exploring the possible range of masses and semimajor-axis of the inner giant planet to form PHPs, we 
  performed histograms taking into account all the simulations in where we found PHPs, without differentiating them according to the size
  of the planetesimals, the type I migration rates, or the gap opening.
  As it is shown in the top left panel in Fig. \ref{fig:fig11}, more than the 70\% of the PHPs were formed in planetary systems with the 
  innermost giant planet located between 3~au and 6~au. Then, following the top right panel of Fig. \ref{fig:fig11}, it seems that a possible range for the 
  mass of the inner giant planet to form PHPs could be between $1\text{M}_{\text{J}}$ and $3\text{M}_{\text{J}}$ since we find more than the
  60\% of the PHPs within these values. These top panels do not distinguish from
 planetary systems with one, two or three gas giants, but the middle and the bottom histograms do. If we distinguish planetary systems with only one 
 gas giant from those with two gas giants, we find some differences. For those systems with one giant, a possible range for the location and mass for the
 giant planet, to form PHPs, is between 3~au and 4~au and between $4\text{M}_{\text{J}}$ and $5\text{M}_{\text{J}}$, respectively. However, for those planetary systems with two gas giants, the majority of the PHPs were formed in systems with the inner giant around 6~au and between $1\text{M}_{\text{J}}$ and $3\text{M}_{\text{J}}$.
   
It is however important to remark that these results are obtained only considering the results of the developed N-body simulations, which do not present a
detailed sample of the semimajor-axis distribution of the inner gas giants between 1.5~au and 10~au. Thus, although the sample of N-body simulations carried out
in this work helps us to explore the ranges of mass and position for the inner giant planet to form PHPs, this sample is not enough to determine an optimal mass and an
  optimal position for the inner giant planet to form PHPs, even more if we want to discriminate between systems with one and two giants. In order to determine a precise value
  of the mass and location of a giant planet to efficiently form PHPs, it is necessary to develop a greater number of N-body simulations from a better and also greater selection
  of initial conditions. Since we count with these initial conditions from the results of PI, we will return to this analysis in future works.  

\section{Contrast with the observed planet population}
\label{sec:sec6}

One of the goals of this work is to contrast the final configurations of our SSAs with those planetary systems already observed
that present at least some similarities with this population. It is, however, important to remember that the goal of PI and this paper is
not to reproduce the current exoplanet population, but to generate a great diversity of planetary systems, particularly SSAs, without 
following any observable distribution in the disc parameters. Then, once the late-accretion stage has finished and the
final configurations of SSAs have been obtained, our goal is to contrast our results with the current planetary systems population 
which present, on the one hand, gas giants like Saturn or Jupiter beyond 1.5~au and, on the other hand, rocky planets in the inner regions,
particularly in their HZs. To do this, we are only going to take into account those planetary systems around solar-type stars with masses in the range of 
$\sim 0.8\text{M}_\odot$ to $1.2\text{M}_\odot$ which present at least one giant planet beyond 1.5 au and those planetary systems which present planets of 
interest near or in the HZ. As we have already mentioned at the beginning, planetary systems similar to our own have not yet been discovered.
However, it is possible that planetary systems with detected gas giants present rocky planets in the inner zone of the disc, and also, it could be possible that 
planetary systems with detected rocky planets present gas giants not yet discovered, particularly if they present large semimajor-axis and 
low eccentricities.

This section is then divided by taking into account the comparison with two different observed planet populations: the PHPs population and the 
gas giant planet population.

\subsection{PHPs population}
\label{subsec:subsec9}

Since the CoRoT \citep{Barge2008} and Kepler \citep{Borucki2010} missions began to work, our conception of the existence of other 
worlds has changed considerably. During their working years, these missions were able to find a very high number of rocky planets with 
similar sizes to that of the Earth, as never before, giving rise to the exploration of this particular type of exoplanets. 
At present, both missions have ceased to function, but there are others on the way that will help to continue the study, detection and 
characterization of rocky planets, such as TESS (Transiting Exoplanet Survey Satellite) and CHEOPS (CHaracterising ExOPlanets Satellite). 
TESS \citep{Ricker2010} is expected to discover thousands of exoplanets in orbit around the brightest stars in the sky through the 
transit method, and CHEOPS \citep{Broeg2013}, which will complement TESS, will characterize already confirmed exoplanets using photometry 
of very high precision to determine the exact radius of planetary bodies of known mass, between $1\text{M}_\oplus$ and $20\text{M}_\oplus$.

Till now, Kepler has discovered a total of 2244 candidates, 2327 confirmed planets and 30 small, less than twice Earth-size, HZ confirmed 
planets around different type of stars, while the K2 mission, which is still working, has discovered 480 new candidates and 292 confirmed 
planets (\url{https://www.nasa.gov/mission_pages/kepler/}). 

\begin{table}
  \centering
  \caption{Potentially habitable planets around solar-mass stars from \url{http://phl.upr.edu}. * planet candidate/unconfirmed. $^*$ estimated value for rocky composition when not available.}
  \label{tab:tab9}
  \begin{tabular}{l|c|c|c|c|c}
    \hline
    Name & Mass [$\text{M}_\oplus$] & Radius [$R_\oplus$] & Spectral Type \\
    \hline
    Kepler-452 b &  $4.7^{*}$     &  1.6                & G \\
    Kepler-22 b  &  $20.4^{*}$    &  2.4                & G \\
    Kepler-69 c  &   $6.0^{*}$    &  1.7               & G \\
    Kepler-439 b &  $19.5^{*}$    &  2.3              & G \\
    tauCet-e*    &  $> 4.3$      &  $1.6^{*}$                & G \\
    HD-40307 g*  &  $> 7.1$      &  $1.8^{*}$               & K \\
    KOI 7235.01* & $1.4^{*}$      & 1.1  & G \\
    KOI 6425.01* & $3.5^{*}$      & 1.5  & G \\
    KOI 7223.01* & $3.7^{*}$ & 1.5  & G \\
    KOI 7179.01* & $1.5^{*}$ & 1.2  & G \\
    KOI 4450.01* & $9.9^{*}$ & 2.0  & G \\
    KOI 4054.01* & $10.1^{*}$ & 2.0 & G\\
    KOI 7587.01* & $14.9^{*}$ & 2.2 & G \\
    KOI 6734.01* & $13.1^{*}$ & 2.1  & G \\
    KOI 7136.01* & $17.4^{*}$ & 2.3 & G \\
    KOI 6676.01* & $7.0^{*}$ & 1.8 & F\\ 
    KOI 5475.01* & $5.0^{*}$ & 1.7  & F\\
    KOI 7554.01* & $9.9^{*}$ & 2.0 & F\\
    KOI 5236.01* & $9.9^{*}$ & 2.0 & F\\
    KOI 4103.01* & $14.4^{*}$ & 2.2  & K\\
    KOI 5202.01* & $7.3^{*}$ & 1.8 & F \\
    KOI 6343.01* & $8.4^{*}$ & 1.9 & F\\
    KOI 7470.01* & $8.4^{*}$ & 1.9 & K\\
    KOI 5276.01* & $15.2^{*}$ & 2.2 & K\\
    KOI 7345.01* & $21.1^{*}$ & 2.4 & F\\
    KOI 4458.01* & $25.9^{*}$ & 2.5 & F\\
    KOI 3946.01* & $20.7^{*}$ & 2.4 & F\\
    KOI 7040.01* & $14.7^{*}$ & 2.2 & F\\   
    \hline
  \end{tabular}
\end{table}

\begin{figure}
  \includegraphics[angle=0, width=\columnwidth]{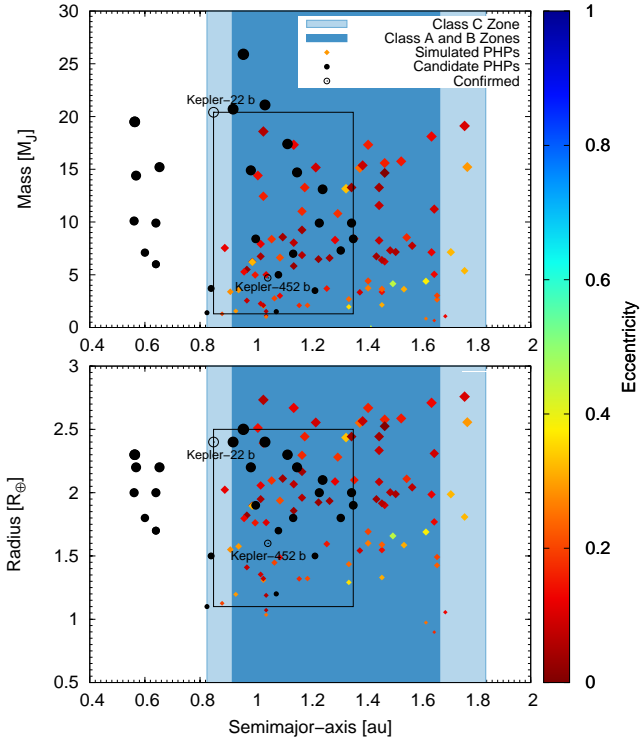}
  \caption{Semimajor-axis vs mass plane for the simulated PHPs population (colored diamonds) and the observed PHPs around solar-mass
stars (black points). The colorscale represents the eccentricity, data that we have for the simulated PHPs but not for the majority of 
the observed ones. The shaded areas represent our definition of the HZ for a $5\text{M}_\oplus$ planet and the black square delimits the region
in where we find both kinds of planets, the observed ones and the simulated ones. Kepler-452 b and Kepler-22 b are the only two confirmed 
exoplanets within these region.}
\label{fig:fig12}
\end{figure}

In order to contrast our simulations with observations, we consider those N-body simulations that, after 200 Myr of evolution, present PHPs.
Then, we contrast the simulated PHPs with the observed ones, only considering those PHPs from \url{http://phl.upr.edu} that
orbit stars with masses in the range of $\sim 0.8\text{M}_\odot$ to $1.2\text{M}_\odot$ (see table \ref{tab:tab9}). It is important to remark that we
  are not making a precise statistical comparison between the observed and the simulated population of PHPs, but trying to determine if there are observed planets
  in the range of masses and radii of the ones we form in our simulations within the HZ. Figure \ref{fig:fig12} shows a semimajor-axis vs mass plane of these
two populations. Most of the masses of the observed exoplanets are estimated considering a rocky composition when not available 
(see \url{http://phl.upr.edu}). The blue and light-blue shaded areas in Fig. \ref{fig:fig12}
represent the class A and B, and C HZ regions, respectively, as we have defined in section \ref{subsec:subsec1}. As we can see in Fig. 
\ref{fig:fig12}, there is an overlap of both populations, the real one and the simulated one, in both panels that represent the mass vs. semimajor-axis and the radius vs. semimajor-axis diagrams.
To plot the radius of our simulated planets we consider physical densities of 5~g~cm$^{-3}$. The black rectangle delimits these regions of overlap between
0.849~au and 1.356~au, and between $1.3\text{M}_\oplus$ and $20.4\text{M}_\oplus$ (panel a) and between 0.849~au and 1.356~au, and between $1.1R_\oplus$ and $2.5R_\oplus$ (panel b),
in where we can find both simulated and observed PHPs. Particularly and till now, the only confirmed exoplanets that remain within this region are Kepler-452 b and 
Kepler-22 b, the rest are still candidates awaiting for confirmation. Kepler-452 b, which presents a radius of 
$1.63^{+0.23}_{-0.20}R_\oplus$ is, until now, the fist planet that presents the longest orbital period for a small transiting exoplanet to date 
and orbits a star of $1.037^{+0.054}_{-0.047}\text{M}_\odot$ at $1.046^{+0.019}_{-0.015}$~au \citep{Jenkins2015}. Kepler-22 b is the only planet 
orbiting a star of $0.970^{+0.06}_{-0.06}\text{M}_\odot$ in the inner edge of its HZ, at $0.849^{+0.018}_{-0.018}$~au, with a radius of 
$2.38^{+0.13}_{-0.13}R_\oplus$, and a mass estimation of $< 36\text{M}_\oplus$ \citep{Borucki2012}. 

It is important to clarify that this comparison we are doing here between the simulated PHPs in some SSAs and the selected observed PHPs that orbit solar-type stars is not a direct comparison since none of these observed PHPs is still part of a planetary system similar to our own, this is,
  none of them has a Saturn or Jupiter analogue associated beyond 1.5~au. Besides, it is worth remarking that the combination of different
  detection techniques that allow us to measure both, the mass and the radius of a planet, have shown that many of the planets discovered with Kepler could
  have compositions very different from those of the terrestrial planets of our Solar system \citep{Wolfgang2016}. Particularly, planets with $M > 6M_\oplus$
  could include a mixture of gas-rich and rocky planets, which in fact could be very different from the PHPs we form within our simulations.
  However we can see that we are able to form PHPs in a region of the semimajor-axis vs mass and semimajor-axis vs radius diagrams which are already explored. 
The lack of PHPs in more extended orbits, this is, until $\sim 2$~au is possible due to the fact that the HZ planets associated with 
G-type main sequence stars have longer orbital periods, which means that more data for the detection is required in comparison with cooler 
K and M-type stars \citep{Batalha2014}. It is then expected that the advances in future missions will help us to detect a greater number of 
PHPs in these regions.

\subsection{Gas giant planet population}
\label{subsec:subsec10}

The population of giant planets, in particular hot-Jupiters, which was at first the most abundant and easy to be detected by both the 
Doppler and the transit techniques, is today, in front of the enormous amount of terrestrial-type exoplanets discovered by Kepler, a rare 
population \citep{Batalha2014}. It is then believed that the rate of occurrence of giant planets is greater beyond 2~au and particularly
around the location of the snowline \citep[see][and references therein]{Fischer2014}. However, it is still not so easy to detect gas giant
planets on wide and circular orbits since \citet{Barnes2007} showed that planets with eccentric orbits are more likely to transit than those
planets with circular orbits with the same semimajor-axis by a factor of $(1-e^2)^{-1}$, where $e$ is the eccentricity of the planet.

To date, $\sim 500$ gas giant planets (more massive than Saturn) orbiting solar-type stars with masses in 
the range of $\sim 0.8\text{M}_\odot$ to $1.2\text{M}_\odot$ were confirmed. From this, only $\sim 142$ are orbiting their star beyond $\sim$ 1.5~au. However
none of these giant planets has yet an associated rocky planet in the inner regions of the system, much less in the HZ.
 This does not strictly mean 
that these rocky companions in the inner zone of the disc do not exist. It is possible that the giants are camouflaging
the signals of the rocky planets and that this makes it more difficult to detect them.
Thus, again, we need to be careful when comparing the giant planets in our simulated planetary systems with those confirmed giants beyond 1.5~au,
since they may still be part of planetary systems very different to Solar system analogues.

\begin{figure*}
  \includegraphics[angle=270, width=\textwidth]{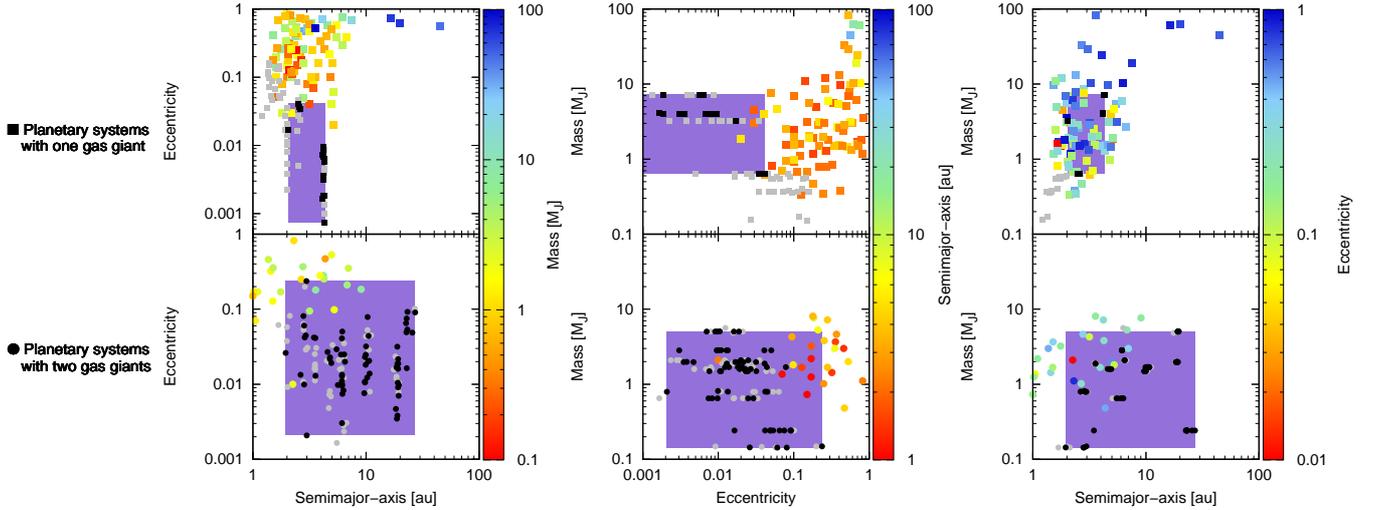}
  \caption{Comparison between the observations and the developed simulations. This plot shows eccentricity vs semimajor-axis,
    mass vs eccentricity and mass vs semimajor-axis planes
    for planetary systems with one gas giant (top panels) and two gas giants (bottom panels). The colored squares and points represent
    the locations of the gas giant planets in planetary systems with only one and two gas giants, respectively, beyond 1.5~au and with
    masses greater than Saturn. The black squares and points represent the locations of the gas giants in simulated planetary systems that
    present PHPs, in planetary systems with only one and two gas giants, respectively. The grey squares and points represent the locations
    of the gas giants in simulated planetary systems that do not present PHPs, in planetary systems with only one and two gas giants,
    respectively. The colored areas represent the range in semimajor-axis, eccentricity and mass, in where we find planetary systems with
    one or two gas giants, that were able to form PHPs.}
\label{fig:fig13}
\end{figure*}

In order to contrast our simulated gas giant population with the observations, we are only going to take into 
account those single or multiple planetary systems, with one or more gas giants with masses greater than $0.3\text{M}_{\text{J}}$ beyond 1.5~au.
We also take into account those multiple planetary systems with more than two gas giants beyond 1~au. 
Figure \ref{fig:fig13} shows semimajor-axis vs eccentricity vs mass planes for 
planetary systems with only one gas giant (top panels) and with two gas giants (bottom panels) 
\citep[\url{http://exoplanet.eu/}][]{Schneider2011}. Each
diagram shows the confirmed gas giant planets with masses greater than Saturn in colored squares 
(planetary systems with only one gas giant) and points (planetary systems with two gas giants) along with the gas giant
planet population of all our simulations. Those gas giants in planetary systems that present PHPs at the end of the N-body simulations
are shown in black squares (planetary systems with only one gas giant) and points (planetary systems with two gas giants), 
while the gas giants in planetary systems that did not form PHPs are represented in grey.

In general, it can be seen that in those planetary systems with only one gas giant (top panels) the range of semimajor-axis and masses
of the simulated gas giants is pretty well represented by the confirmed population of giant planets. However,
this does not seem to happen for the eccentricity range. The least eccentric confirmed exoplanet of the selected sample has an eccentricity
of 0.02, whereas we form planets with up to more than 
one order of magnitude less eccentric. This population of planetary systems with very low eccentric gas giants seems to be not yet 
detected, and in particular, it seems to be the one that is able to form PHPs. It is also interesting to remark that those simulated gas giants with
eccentricities higher than 0.2, semimajor-axis greater than $\sim$ 1.5~au and masses greater than $\sim 0.3\text{M}_{\text{J}}$ and which are overlapped
with the observed and confirmed gas giants, are part of planetary systems that were not efficient in forming PHPs (grey squares).

Something similar seems to happen in planetary systems with two gas giant planets (bottom panels). In this kind of
planetary systems, although the range in mass is well represented by the confirmed population of giant planets, the ranges
of semimajor-axis and eccentricities are not. In this sense, we are forming a population of planetary systems with very low eccentric gas giants in 
wider orbits extended until $\sim$30~au that seems to be not yet detected. And again, this population seems to be efficient in forming PHPs.

We then ask ourselves if in fact this population of very low eccentricities and wide orbits is a natural result of the evolution of the 
simulations, and therefore is, indeed, a population not yet detected, or if it is a consequence of the initial conditions imposed by our 
model of planet formation, where at the beginning and at the end of the gaseous stage the planets are in circular and co-planar orbits. 
A detailed study of the evolution of the semimajor-axis and eccentricities of our giants shows us that the gravitational interactions 
between these bodies and the rest of the planets of the system, which are consider during the evolution of the N-body simulations,
lead to significant changes in their eccentricities although do not lead to planet-planet 
scattering events that involve gas giant planets. For planetary systems with only one giant planet, the registered 
changes in the giants eccentricities during the late-accretion stage reach values higher than an order 
of magnitude of the initial eccentricities. For planetary systems with two gas giants these changes are even greater, and the eccentricities can 
increase even more. Then, this population of planetary systems with giant planets of low eccentricities is, within the framework of having
  considered initial (this is at the beginning of the N-body simulations) quasi-circular and co-planar orbits for the giants, a natural result of
  its own orbital evolution. It is however important
to highlight that more realistic treatments for the gaseous phase evolution of these systems could lead to different gas giant planet
configurations. Our model of planet formation for this stage does not include the gravitational interactions between
planets. These interactions could lead to encounters between giants increasing their eccentricities as the gas dissipates and scattering other
planets in their way. On the other hand, this effect could also cause planets to be trapped in mean motion resonances. Both situations can in turn lead to different
final configurations for the planetary systems at the end of the late accretion stage.

\section{Discussion}
\label{sec:sec7}

In the present work, we carry out N-body simulations aimed at studying the late accretion in Solar system analogues (SSAs), and the formation
of potentially habitable planets and water delivery. To interpret the results of our simulations it is very important to mention some points
concerning the evolution of our systems during the gaseous phase as well as once the gas has dissipated.

According to that mentioned in Section \ref{sec:sec3}, we select 10 scenarios of work in order to carry out our analysis concerning the late-accretion
stage in SSAs. The evolution of such systems during the gaseous phase was obtained from the work developed by \citet{Ronco2017}, PI.

It is worth noting that, as it was mentioned in Section \ref{sec:sec2}, the model of planet formation of PI, from which we obtain
  the initial conditions to develope N-body simulations, presents limitations. Some of these limitations are the consideration of isothermal discs and the
  corresponding type I migration rates for such discs, and the lack of the calculation of the gravitational interactions between embryos. The
  consideration of more complex physics during the gaseous phase could alter the final results of embryo and planetesimal distributions at
  the end of the gaseous phase and thus significantly change our results on the final planet mass, their
  locations and water contents at the end of the late-accretion stages.

  Particularly, as it was mentioned before, type I migration rates for non-isothermal discs could be very different for those in isothermal discs such as they
  could stop or even change the migration direction of the planets \citep{Paardekooper2010,Paardekooper2011}.
  The consideration of gravitational interactions between planets during the gaseous phase could 
  allowed the planets to be trapped in mean motion resonances (MMR) and this could avoid a fast orbital decay into very inner zones of the disc
  \citep{MassetSnellgrove2001}. Moreover, the consideration of this effect, as it was shown by \citet{MorbidelliCrida2007} within the framework
  of our Solar system is that, if Jupiter was able to open a gap in the disc and migrate inward through type II migration, and if at the same
  time Saturn was able to migrate faster towards Jupiter, they could have been locked in a 2:3 (MMR) and could have stopped or even reversed the
  migration of both planets together. On the other hand, the interactions between giant planets in the inner regions of the disc when the gas is already
  removed due to photoevaporation, can lead to the ejection of one or two planets exalting the eccentricities of the remaining ones.
  Then, all these considerations, which were not taken into account in PI, can change the final results of our N-body simulations.
  
It is also important to remark that the 10 scenarios of work chosen to carry out our investigation show differences concerning the number
and the physical and orbital properties of the giant planets that host. On the one hand, 5 of those 10 scenarios start the post-gas phase
with a single giant planet, whose mass ranges from $0.45\text{M}_{\text{S}}$ to $4\text{M}_{\text{J}}$. On the other hand, the other five systems of work
host two gaseous giants at the beginning of the post-gas stage, most of which show masses of the order of the Jupiter's mass.

In those 5 scenarios of work that host 2 gaseous giants, the initial orbital separation between them ranges between 4.1 and 11.7 mutual
Hill radii. It is worth noting that none of the N-body simulations developed in the present research produced close encounters between the
massive giant planets of the system, which seems to be consistent with the works developed by \citet{Gladman1993}, \citet{Chambers1996}, and
\citet{Marzari2014}. Obviously, an analysis concerning the stability of the system of 2 giant planets is more complex since they are immersed
in a population of embryos and planetesimals in our scenarios of study. However, this simple consideration allows us to understand the absence
of close encounters between the massive giant planets in all our N-body simulations.

From this, we must remark that the SSAs formed in the present research did not experiment resonance capture processes during the gaseous
phase while they did not undergo strong dynamical instability events that involve the massive giant planets. Thus, the proposed analogy between
the systems formed in our simulations and the Solar system should be carefully interpreted if scenarios of formation such as the Grand Tack
model \citep{Walsh2011} and the Nice model \citep{Tsiganis2005} are accepted. Studies concerning the formation and evolution of systems
that host giant planets and undergo strong instability events between them will be addressed in future works.

On the other hand, it is interesting to note that our rocky planets grow fundamentally from the giant impacts they undergo
once the gas disk dissipates, a process that lasts about 100 Myr. Late formation stages like the ``late veneer'' or the ``heavy bombardment''
on our Earth are not present in our simulations. If our Solar system experienced three stages of formation given by the stage of giant 
impacts, the late-veneer, and finally the heavy bombardment \citep[see][and references therein]{Raymond2013}, the planets of our 
simulated systems only suffer the first stage, not being affected by later ones in which a population of planetesimal remnants reshape them.

Our N-body simulations form, after 200 Myr of evolution, planetary systems similar to our own with rocky 
planets in the inner zone of the disc and at least one gas giant beyond 1.5~au. One of the goals of this work is to analyze the formation
of planets within the HZ. The simulations performed were able to form two 
very different kinds of PHPs: dry and water-rich (this is, planets that come from beyond the snowline or planets that have accreted at least an embryo from this region).
Taking into account 
the 4 default scenarios described in this work, the most common type of PHPs seems to be the dry ones since we formed them in 3 
of those 4 scenarios. However, considering the final number of
PHPs formed within all the developed simulations, and considering that the most common result from PI was to form SSAs in scenarios with
planetesimals of 100~m, it seems that the most common PHPs are the water-rich ones instead of
the dry ones, since a 65\% of the whole PHPs population is formed by water-rich planets (almost all of them formed in scenarios with 
planetesimals of 100~m), and a 35\% is formed by dry planets.

One of the limitations of our model, which is important to remark, is that the high water contents in those water-rich PHPs must be considered as upper 
limits due to the fact that we do not take into account water loss during impacts. However, the great majority of our water-rich PHPs 
do not suffer from a great number of giant collisions. Thus, their primordial water amounts could survive during the whole evolution. 
Moreover, \citet{Dvorak2015} showed that the water retained in fragments after a collision remains on the target for impact velocities 
$v \leq 1.3v_{\text{esc}}$, while more than 80\% of the water stays as well in strongly inclined hit-and-run scenarios.
Another interesting result is that, due to the fact that all our water-rich PHPs come from beyond the snowline, their water contents
are primordial. This result is in concordance with the results of \citet{Zain2017} who also found that all water contents 
of the HZ planets in systems that harbor a Jupiter-mass planet around the snowline, are also primordial.  

If, although the water contents have to be considered as upper limits, they could have remained in their great totality until the end 
of the evolution. We wonder then what is the effect of such water contents in the habitability. 
Some authors argue that too large amounts of water may be detrimental for the potential habitability 
\citep{Alibert2014,Kitzmann2015}. These authors claim that a too large water layer can lead to the existence of a high 
pressure ice layer at the bottom of the global ocean, and this could prevent the Carbonate-silicate cycle from occurring. This
could be a problem since the Carbonate-silicate cycle is fundamental to maintain the temperature of the surface for long periods of
time \citep{Alibert2014}, and could be particularly problematic for stars such as the sun, which suffer variations in their flux at 
relatively short timescales. Moreover, \citet{Kitzmann2015} showed that an unstable CO$_2$ cycle could
limit the extension of the HZ for ocean planets. Other studies, such as the work developed by \citet{Abbot2012}, 
found that the weathering behavior is not sensitive to the land fraction for those planets which are covered by oceans, 
as long as the land fraction is greater than 0.01. This result suggests that those partially covered water planets may have a HZ 
similar to a planet which is not. Moreover, although these authors indicate that water worlds could have narrower HZ, they 
could be capable of ``self-arresting'' a moist greenhouse effect, and could become planets with partial ocean coverage. Recently,
\citet{Noack2016} also concluded that water-rich planets may be habitable and could allow the developing of live as we know it, if
these water-rich planets present shallow oceans, a low planet mass, or a high surface temperature. They find that small-massive planets
with less than one Earth mass can be habitable even for substantial amounts of water. However, Super-Earths covered with oceans can only
be considered as habitable planets, if they contain low percentages of water by mass.
Although this is a topic that is still under debate and that the only existence of liquid water is definitely not enough to determine
habitability, water-rich planets seem to be common in the Universe \citep{Raymond2004,Simpson2017}. We particularly find them in planetary 
systems without gas giant planets in high \citep{deElia2013} and low-mass discs \citep{Roncodeelia2014,Ronco2015}, in planetary systems
around low-mass stars \citep{Dugaro2016}, and in planetary systems with different gas giant planets as main perturbers \citep{Zain2017},
and because of that we still consider they present a particular interest.

Another important topic that could affect the habitability of our PHPs due to the possible absence of plate tectonics is their final 
masses, since the great majority of them are Super-Earths (59\%) and, to a lesser extent, Mega-Earths (31\%), with masses very different 
from that of our planet. However, \citet{Valencia2007} show that plate tectonics, which are a necessary condition for the development 
of life, can also be found in Super-Earths. These authors show that, as the planetary mass increases, the subduction process, and therefore the 
plate tectonics, becomes easier. Therefore, massive Super-Earths are very likely to exhibit plate tectonics and therefore, to keep life 
possible. On the other hand, \citet{Morard2011} showed that, for masses greater than $2\text{M}_\oplus$, the Super-Earths cores could be entirely solid,
this precludes the existence of a liquid metallic core driven by magnetic fields. The lack of magnetic fields is detrimental to the potential
habitability given that they are those that favor the existence and permanence of an atmosphere for long periods of time.

Finally, and following with our model limitations, we believe that a more realistic treatment of the collisions, including fragmentation and
hit-and-run collisions \citep{Chambers2013} rather than assuming that all impacts lead to a perfect merger of the colliding bodies, should 
be included in future works. \citet{Quintana2016} presented results of N-body simulations of the late stage formation of Earth-like planets
around sun-like stars with Jupiter and Saturn analogues perturbing the system and including a collision model that follows \citet{LeinhardtStewart2012} and
\citet{Chambers2013}. They found that, globally, when fragmentation is considered the number of final planets and their masses are similar
to those formed in the standard model without fragmentation and hit-and-run collisions. However, the collision history of the final planets is
different and the accretion timescales are usually doubled in the fragmentation simulations. Thus, we consider that this improvements in the treatment
of collisions will allow us to determine in more detail the orbital and physical properties of the planets formed in our simulations, particularly the
final masses and amounts of water by mass of the formed PHPs.

\section{Conclusions}
\label{sec:sec8}
In this work we carry out a great number of N-body simulations aimed at analyzing the global formation of
the post-gas phase of SSAs, particularly focusing in the formation of rocky planets within the HZ and their final water contents.
The initial conditions to start the simulations were provided by the results of PI \citep{Ronco2017}. In that first work we analyzed 
the evolution, during the gas-phase, of different planetary systems, particularly of those which were at the end of the gaseous phase 
similar to our own, considering different planetesimal sizes and different type I migration rates.

The main results of our simulations, which are analyzed for planetary systems that did not experienced type I migration rates during
the gaseous phase, show that the efficiency in the formation of potentially habitable planets (PHPs) seems to
strongly depend on the size of the planetesimals from which these systems were formed. Planetary systems formed from small
planetesimals are the most efficient ones. Moreover, this kind of planetary systems, formed from small planetesimals of 100~m, 
is the only one capable of forming water-rich PHPs. Bigger planetesimals of 1~km and 10~km also form PHPs but completely dry, 
and planetary systems formed from planetesimals of 100~km do not form PHPs at all.

We also test the sensitivity of the results to scenarios that experienced type I migration rates during the gaseous phase and 
that were also able to form giant planets massive enough to open a gap in the gas disc and to \emph{turn on} type II migration.
Our results show that planetary systems which experienced very low type I migration rates do not present significant differences to the
results of those SSAs that did not, and that planetary systems which experienced type I migration rated reduced to a 10\% seems to favor the 
formation of PHPs in scenarios formed from small planetesimals but is detrimental for the formation of PHPs in scenarios formed
from big planetesimals. Those planetary systems which were able to form giant planets that opened gaps in the gas discs during
the gaseous phase seem to act in the same way as type I migration does. Analyzing all the developed simulations, it also seems that 
there might be an optimal mass and location for the inner giant planet of the system regarding the efficiency of forming PHPs. Although 
we estimate those values around $\sim$ 6~au and between $1\text{M}_{\text{J}}$ and $2\text{M}_{\text{J}}$ taking into account all the developed simulations in this
work, a most detailed set of numerical simulations considering a the semimajor-axis distribution of the inner gas giants between 1.5
au and 10 au should be developed in order to properly determine these values. It is however important to remark that we count with all the information
to do this work in a near future.

Finally, we make a comparison between our simulated planetary systems and the current potentially habitable planet population and gas giant
population. Although this kind of comparisons must be analyzed carefully since, to date, no planetary systems 
similar to our own have been yet discovered, this is, none of the observed (confirmed and candidates) potentially habitable planets present
a gas giant companion beyond 1.5~au, we are able to form PHPs in a region of the mass vs semimajor-axis diagram that is currently 
being explored. Moreover, the simulated giant planet population that was capable of forming PHPs occupies a region of the semimajor-axis vs 
eccentricity diagram of extended semimajor-axis and very low eccentricities, and to date a population with these characteristics has not 
been yet discovered. We believe that the discovery of exoplanets with these properties is a challenge of future missions. Particularly,
\citet{Hippke2015} explored the potential of future missions such as PLATO 2.0 in detecting Solar system analogues, and showed that the
discovery of Venus and Earth-like planets transiting stars like our sun is feasible at high signal-to-noise ratio after
collecting 6 yrs of data. They also showed that, due to the high number of observed stars by PLATO, the detection of single-transit events by
cold gas giants like Jupiter, Saturn, Uranus, and Neptune analogues, will be possible. These future detections, particularly those of cold gas giants,
will allow us to associate this population with the potential existence of planets in the HZ.

\section*{Acknowledgements}
The authors thank the anonymous referee for her or his valuable comments,
which helped to improve the manuscript. M.P.R. thanks Octavio Miguel Guilera for useful discussions. This work has been supported by grants from the
National Scientific and Technical Research Council through the PIP 0436/13, and National
University of La Plata, Argentina, through the PIDT 11/G144.  




\bibliographystyle{mnras}
\bibliography{Bibliography} 

\begin{thebibliography}{}
\makeatletter
\relax
\def\mn@urlcharsother{\let\do\@makeother \do\$\do\&\do\#\do\^\do\_\do\%\do\~}
\def\mn@doi{\begingroup\mn@urlcharsother \@ifnextchar [ {\mn@doi@}
  {\mn@doi@[]}}
\def\mn@doi@[#1]#2{\def\@tempa{#1}\ifx\@tempa\@empty \href
  {http://dx.doi.org/#2} {doi:#2}\else \href {http://dx.doi.org/#2} {#1}\fi
  \endgroup}
\def\mn@eprint#1#2{\mn@eprint@#1:#2::\@nil}
\def\mn@eprint@arXiv#1{\href {http://arxiv.org/abs/#1} {{\tt arXiv:#1}}}
\def\mn@eprint@dblp#1{\href {http://dblp.uni-trier.de/rec/bibtex/#1.xml}
  {dblp:#1}}
\def\mn@eprint@#1:#2:#3:#4\@nil{\def\@tempa {#1}\def\@tempb {#2}\def\@tempc
  {#3}\ifx \@tempc \@empty \let \@tempc \@tempb \let \@tempb \@tempa \fi \ifx
  \@tempb \@empty \def\@tempb {arXiv}\fi \@ifundefined
  {mn@eprint@\@tempb}{\@tempb:\@tempc}{\expandafter \expandafter \csname
  mn@eprint@\@tempb\endcsname \expandafter{\@tempc}}}

\bibitem[\protect\citeauthoryear{{Abbot}, {Cowan}  \& {Ciesla}}{{Abbot}
  et~al.}{2012}]{Abbot2012}
{Abbot} D.~S.,  {Cowan} N.~B.,   {Ciesla} F.~J.,  2012, \mn@doi [\apj]
  {10.1088/0004-637X/756/2/178}, \href
  {http://adsabs.harvard.edu/abs/2012ApJ...756..178A} {756, 178}

\bibitem[\protect\citeauthoryear{{Abe}, {Ohtani}, {Okuchi}, {Righter}  \&
  {Drake}}{{Abe} et~al.}{2000}]{Abe-2000}
{Abe} Y.,  {Ohtani} E.,  {Okuchi} T.,  {Righter} K.,   {Drake} M.,  2000,
  {Water in the Early Earth}.
pp 413--433

\bibitem[\protect\citeauthoryear{{Alibert}}{{Alibert}}{2014}]{Alibert2014}
{Alibert} Y.,  2014, \mn@doi [\aap] {10.1051/0004-6361/201322293}, \href
  {http://adsabs.harvard.edu/abs/2014A%26A...561A..41A} {561, A41}

\bibitem[\protect\citeauthoryear{{Alibert}, {Mordasini}, {Benz}  \&
  {Winisdoerffer}}{{Alibert} et~al.}{2005}]{Alibert2005}
{Alibert} Y.,  {Mordasini} C.,  {Benz} W.,   {Winisdoerffer} C.,  2005, \mn@doi
  [\aap] {10.1051/0004-6361:20042032}, \href
  {http://adsabs.harvard.edu/abs/2005A%26A...434..343A} {434, 343}

\bibitem[\protect\citeauthoryear{{Andrews}, {Wilner}, {Hughes}, {Qi}  \&
  {Dullemond}}{{Andrews} et~al.}{2010}]{Andrews2010}
{Andrews} S.~M.,  {Wilner} D.~J.,  {Hughes} A.~M.,  {Qi} C.,   {Dullemond}
  C.~P.,  2010, \mn@doi [\apj] {10.1088/0004-637X/723/2/1241}, \href
  {http://adsabs.harvard.edu/abs/2010ApJ...723.1241A} {723, 1241}

\bibitem[\protect\citeauthoryear{{Barge} et~al.,}{{Barge}
  et~al.}{2008}]{Barge2008}
{Barge} P.,  et~al., 2008, \mn@doi [\aap] {10.1051/0004-6361:200809353}, \href
  {http://adsabs.harvard.edu/abs/2008A%26A...482L..17B} {482, L17}

\bibitem[\protect\citeauthoryear{{Barnes}}{{Barnes}}{2007}]{Barnes2007}
{Barnes} J.~W.,  2007, \mn@doi [\pasp] {10.1086/522039}, \href
  {http://adsabs.harvard.edu/abs/2007PASP..119..986B} {119, 986}

\bibitem[\protect\citeauthoryear{{Batalha}}{{Batalha}}{2014}]{Batalha2014}
{Batalha} N.~M.,  2014, \mn@doi [Proceedings of the National Academy of
  Science] {10.1073/pnas.1304196111}, \href
  {http://adsabs.harvard.edu/abs/2014PNAS..11112647B} {111, 12647}

\bibitem[\protect\citeauthoryear{{Bitsch}, {Johansen}, {Lambrechts}  \&
  {Morbidelli}}{{Bitsch} et~al.}{2015}]{Bistch2015}
{Bitsch} B.,  {Johansen} A.,  {Lambrechts} M.,   {Morbidelli} A.,  2015,
  \mn@doi [\aap] {10.1051/0004-6361/201424964}, \href
  {http://adsabs.harvard.edu/abs/2015A%26A...575A..28B} {575, A28}

\bibitem[\protect\citeauthoryear{{Bolmont}, {Libert}, {Leconte}  \&
  {Selsis}}{{Bolmont} et~al.}{2016}]{Bolmont2016}
{Bolmont} E.,  {Libert} A.-S.,  {Leconte} J.,   {Selsis} F.,  2016, \mn@doi
  [\aap] {10.1051/0004-6361/201628073}, \href
  {http://adsabs.harvard.edu/abs/2016A%26A...591A.106B} {591, A106}

\bibitem[\protect\citeauthoryear{{Borucki} et~al.,}{{Borucki}
  et~al.}{2010}]{Borucki2010}
{Borucki} W.~J.,  et~al., 2010, \mn@doi [Science] {10.1126/science.1185402},
  \href {http://adsabs.harvard.edu/abs/2010Sci...327..977B} {327, 977}

\bibitem[\protect\citeauthoryear{{Borucki} et~al.,}{{Borucki}
  et~al.}{2012}]{Borucki2012}
{Borucki} W.~J.,  et~al., 2012, \mn@doi [\apj] {10.1088/0004-637X/745/2/120},
  \href {http://adsabs.harvard.edu/abs/2012ApJ...745..120B} {745, 120}

\bibitem[\protect\citeauthoryear{{Broeg} et~al.,}{{Broeg}
  et~al.}{2013}]{Broeg2013}
{Broeg} C.,  et~al., 2013, in European Physical Journal Web of Conferences. p.
  03005 (\mn@eprint {arXiv} {1305.2270}), \mn@doi{10.1051/epjconf/20134703005}

\bibitem[\protect\citeauthoryear{{Chambers}}{{Chambers}}{1999}]{Chambers1999}
{Chambers} J.~E.,  1999, \mn@doi [\mnras] {10.1046/j.1365-8711.1999.02379.x},
  \href {http://adsabs.harvard.edu/abs/1999MNRAS.304..793C} {304, 793}

\bibitem[\protect\citeauthoryear{{Chambers}}{{Chambers}}{2001}]{Chambers2001}
{Chambers} J.~E.,  2001, \mn@doi [\icarus] {10.1006/icar.2001.6639}, \href
  {http://adsabs.harvard.edu/abs/2001Icar..152..205C} {152, 205}

\bibitem[\protect\citeauthoryear{{Chambers}}{{Chambers}}{2008}]{Chambers2008}
{Chambers} J.,  2008, \mn@doi [\icarus] {10.1016/j.icarus.2008.06.011}, \href
  {http://adsabs.harvard.edu/abs/2008Icar..198..256C} {198, 256}

\bibitem[\protect\citeauthoryear{{Chambers}}{{Chambers}}{2013}]{Chambers2013}
{Chambers} J.~E.,  2013, \mn@doi [\icarus] {10.1016/j.icarus.2013.02.015},
  \href {http://adsabs.harvard.edu/abs/2013Icar..224...43C} {224, 43}

\bibitem[\protect\citeauthoryear{{Chambers}, {Wetherill}  \& {Boss}}{{Chambers}
  et~al.}{1996}]{Chambers1996}
{Chambers} J.~E.,  {Wetherill} G.~W.,   {Boss} A.~P.,  1996, \mn@doi [\icarus]
  {10.1006/icar.1996.0019}, \href
  {http://adsabs.harvard.edu/abs/1996Icar..119..261C} {119, 261}

\bibitem[\protect\citeauthoryear{{Crida}, {Morbidelli}  \& {Masset}}{{Crida}
  et~al.}{2006}]{Crida2006}
{Crida} A.,  {Morbidelli} A.,   {Masset} F.,  2006, \mn@doi [\icarus]
  {10.1016/j.icarus.2005.10.007}, \href
  {http://adsabs.harvard.edu/abs/2006Icar..181..587C} {181, 587}

\bibitem[\protect\citeauthoryear{{D'Angelo} \& {Marzari}}{{D'Angelo} \&
  {Marzari}}{2012}]{DangeloMarzari2012}
{D'Angelo} G.,  {Marzari} F.,  2012, \mn@doi [\apj]
  {10.1088/0004-637X/757/1/50}, \href
  {http://adsabs.harvard.edu/abs/2012ApJ...757...50D} {757, 50}

\bibitem[\protect\citeauthoryear{{Dauphas} \& {Pourmand}}{{Dauphas} \&
  {Pourmand}}{2011}]{Dauphas2011}
{Dauphas} N.,  {Pourmand} A.,  2011, \mn@doi [\nat] {10.1038/nature10077},
  \href {http://adsabs.harvard.edu/abs/2011Natur.473..489D} {473, 489}

\bibitem[\protect\citeauthoryear{{Dugaro}, {de El{\'{\i}}a}, {Brunini}  \&
  {Guilera}}{{Dugaro} et~al.}{2016}]{Dugaro2016}
{Dugaro} A.,  {de El{\'{\i}}a} G.~C.,  {Brunini} A.,   {Guilera} O.~M.,  2016,
  \mn@doi [\aap] {10.1051/0004-6361/201628355}, \href
  {http://adsabs.harvard.edu/abs/2016A%26A...596A..54D} {596, A54}

\bibitem[\protect\citeauthoryear{{Dvorak}, {Maindl}, {Burger}, {Sch{\"a}fer}
  \& {Speith}}{{Dvorak} et~al.}{2015}]{Dvorak2015}
{Dvorak} R.,  {Maindl} T.~I.,  {Burger} C.,  {Sch{\"a}fer} C.,   {Speith} R.,
  2015, Nonlinear Phenomena in Complex Systems, Vol.18, No.3, pp.~310-325,
  \href {http://adsabs.harvard.edu/abs/2015NPCS...18..310D} {18, 310}

\bibitem[\protect\citeauthoryear{{Fischer}, {Howard}, {Laughlin}, {Macintosh},
  {Mahadevan}, {Sahlmann}  \& {Yee}}{{Fischer} et~al.}{2014}]{Fischer2014}
{Fischer} D.~A.,  {Howard} A.~W.,  {Laughlin} G.~P.,  {Macintosh} B.,
  {Mahadevan} S.,  {Sahlmann} J.,   {Yee} J.~C.,  2014, \mn@doi [Protostars and
  Planets VI] {10.2458/azu_uapress_9780816531240-ch031}, \href
  {http://adsabs.harvard.edu/abs/2014prpl.conf..715F} {pp 715--737}

\bibitem[\protect\citeauthoryear{{Fogg} \& {Nelson}}{{Fogg} \&
  {Nelson}}{2005}]{FoggNelson2005}
{Fogg} M.~J.,  {Nelson} R.~P.,  2005, \mn@doi [\aap]
  {10.1051/0004-6361:20053453}, \href
  {http://adsabs.harvard.edu/abs/2005A%26A...441..791F} {441, 791}

\bibitem[\protect\citeauthoryear{{Fogg} \& {Nelson}}{{Fogg} \&
  {Nelson}}{2006}]{FoggNelson2006}
{Fogg} M.~J.,  {Nelson} R.~P.,  2006, \mn@doi [International Journal of
  Astrobiology] {10.1017/S1473550406003016}, \href
  {http://adsabs.harvard.edu/abs/2006IJAsB...5..199F} {5, 199}

\bibitem[\protect\citeauthoryear{{Fogg} \& {Nelson}}{{Fogg} \&
  {Nelson}}{2007}]{FoggNelson2007a}
{Fogg} M.~J.,  {Nelson} R.~P.,  2007, \mn@doi [\aap]
  {10.1051/0004-6361:20066171}, \href
  {http://adsabs.harvard.edu/abs/2007A%26A...461.1195F} {461, 1195}

\bibitem[\protect\citeauthoryear{{Fogg} \& {Nelson}}{{Fogg} \&
  {Nelson}}{2009}]{FoggNelson2009}
{Fogg} M.~J.,  {Nelson} R.~P.,  2009, \mn@doi [\aap]
  {10.1051/0004-6361/200811305}, \href
  {http://adsabs.harvard.edu/abs/2009A%26A...498..575F} {498, 575}

\bibitem[\protect\citeauthoryear{{Gladman}}{{Gladman}}{1993}]{Gladman1993}
{Gladman} B.,  1993, \mn@doi [\icarus] {10.1006/icar.1993.1169}, \href
  {http://adsabs.harvard.edu/abs/1993Icar..106..247G} {106, 247}

\bibitem[\protect\citeauthoryear{{Guilera} \& {S{\'a}ndor}}{{Guilera} \&
  {S{\'a}ndor}}{2017}]{GuileraSandor2017}
{Guilera} O.~M.,  {S{\'a}ndor} Z.,  2017, \mn@doi [\aap]
  {10.1051/0004-6361/201629843}, \href
  {http://adsabs.harvard.edu/abs/2017A%26A...604A..10G} {604, A10}

\bibitem[\protect\citeauthoryear{{Guilera}, {Brunini}  \&
  {Benvenuto}}{{Guilera} et~al.}{2010}]{Guilera2010}
{Guilera} O.~M.,  {Brunini} A.,   {Benvenuto} O.~G.,  2010, \mn@doi [\aap]
  {10.1051/0004-6361/201014365}, \href
  {http://adsabs.harvard.edu/abs/2010A%26A...521A..50G} {521, A50}

\bibitem[\protect\citeauthoryear{{Guilera}, {Fortier}, {Brunini}  \&
  {Benvenuto}}{{Guilera} et~al.}{2011}]{Guilera2011}
{Guilera} O.~M.,  {Fortier} A.,  {Brunini} A.,   {Benvenuto} O.~G.,  2011,
  \mn@doi [\aap] {10.1051/0004-6361/201015731}, \href
  {http://adsabs.harvard.edu/abs/2011A%26A...532A.142G} {532, A142}

\bibitem[\protect\citeauthoryear{{Guilera}, {de El{\'{\i}}a}, {Brunini}  \&
  {Santamar{\'{\i}}a}}{{Guilera} et~al.}{2014}]{Guilera2014}
{Guilera} O.~M.,  {de El{\'{\i}}a} G.~C.,  {Brunini} A.,   {Santamar{\'{\i}}a}
  P.~J.,  2014, \mn@doi [\aap] {10.1051/0004-6361/201322061}, \href
  {http://adsabs.harvard.edu/abs/2014A%26A...565A..96G} {565, A96}

\bibitem[\protect\citeauthoryear{{Hayashi}}{{Hayashi}}{1981}]{Hayashi1981}
{Hayashi} C.,  1981, \mn@doi [Progress of Theoretical Physics Supplement]
  {10.1143/PTPS.70.35}, \href
  {http://adsabs.harvard.edu/abs/1981PThPS..70...35H} {70, 35}

\bibitem[\protect\citeauthoryear{{Hellary} \& {Nelson}}{{Hellary} \&
  {Nelson}}{2012}]{HellaryNelson2012}
{Hellary} P.,  {Nelson} R.~P.,  2012, \mn@doi [\mnras]
  {10.1111/j.1365-2966.2011.19815.x}, \href
  {http://adsabs.harvard.edu/abs/2012MNRAS.419.2737H} {419, 2737}

\bibitem[\protect\citeauthoryear{{Higuchi}, {Kokubo}, {Kinoshita}  \&
  {Mukai}}{{Higuchi} et~al.}{2007}]{HiguchiKokubo2007}
{Higuchi} A.,  {Kokubo} E.,  {Kinoshita} H.,   {Mukai} T.,  2007, \mn@doi [\aj]
  {10.1086/521815}, \href {http://adsabs.harvard.edu/abs/2007AJ....134.1693H}
  {134, 1693}

\bibitem[\protect\citeauthoryear{{Hippke} \& {Angerhausen}}{{Hippke} \&
  {Angerhausen}}{2015}]{Hippke2015}
{Hippke} M.,  {Angerhausen} D.,  2015, \mn@doi [\apj]
  {10.1088/0004-637X/810/1/29}, \href
  {http://adsabs.harvard.edu/abs/2015ApJ...810...29H} {810, 29}

\bibitem[\protect\citeauthoryear{{Ida} \& {Lin}}{{Ida} \&
  {Lin}}{2004}]{IdaLin2004a}
{Ida} S.,  {Lin} D.~N.~C.,  2004, \mn@doi [\apj] {10.1086/381724}, \href
  {http://adsabs.harvard.edu/abs/2004ApJ...604..388I} {604, 388}

\bibitem[\protect\citeauthoryear{{Inaba}, {Tanaka}, {Nakazawa}, {Wetherill}  \&
  {Kokubo}}{{Inaba} et~al.}{2001}]{Inaba2001}
{Inaba} S.,  {Tanaka} H.,  {Nakazawa} K.,  {Wetherill} G.~W.,   {Kokubo} E.,
  2001, \mn@doi [\icarus] {10.1006/icar.2000.6533}, \href
  {http://adsabs.harvard.edu/abs/2001Icar..149..235I} {149, 235}

\bibitem[\protect\citeauthoryear{{Inamdar} \& {Schlichting}}{{Inamdar} \&
  {Schlichting}}{2015}]{InamdarSchlichting2015}
{Inamdar} N.~K.,  {Schlichting} H.~E.,  2015, \mn@doi [\mnras]
  {10.1093/mnras/stv030}, \href
  {http://adsabs.harvard.edu/abs/2015MNRAS.448.1751I} {448, 1751}

\bibitem[\protect\citeauthoryear{{Jacobson}, {Morbidelli}, {Raymond},
  {O'Brien}, {Walsh}  \& {Rubie}}{{Jacobson} et~al.}{2014}]{Jacobson2014}
{Jacobson} S.~A.,  {Morbidelli} A.,  {Raymond} S.~N.,  {O'Brien} D.~P.,
  {Walsh} K.~J.,   {Rubie} D.~C.,  2014, \mn@doi [\nat] {10.1038/nature13172},
  \href {http://adsabs.harvard.edu/abs/2014Natur.508...84J} {508, 84}

\bibitem[\protect\citeauthoryear{{Jenkins} et~al.,}{{Jenkins}
  et~al.}{2015}]{Jenkins2015}
{Jenkins} J.~M.,  et~al., 2015, \mn@doi [\aj] {10.1088/0004-6256/150/2/56},
  \href {http://adsabs.harvard.edu/abs/2015AJ....150...56J} {150, 56}

\bibitem[\protect\citeauthoryear{{Kasting}, {Whitmire}  \&
  {Reynolds}}{{Kasting} et~al.}{1993}]{Kasting1993}
{Kasting} J.~F.,  {Whitmire} D.~P.,   {Reynolds} R.~T.,  1993, \mn@doi
  [\icarus] {10.1006/icar.1993.1010}, \href
  {http://adsabs.harvard.edu/abs/1993Icar..101..108K} {101, 108}

\bibitem[\protect\citeauthoryear{{Kitzmann} et~al.,}{{Kitzmann}
  et~al.}{2015}]{Kitzmann2015}
{Kitzmann} D.,  et~al., 2015, ArXiv e-prints 1507.01727, \href
  {http://adsabs.harvard.edu/abs/2015arXiv150701727K} {}

\bibitem[\protect\citeauthoryear{{Kopparapu} et~al.,}{{Kopparapu}
  et~al.}{2013a}]{Kopparapu2013}
{Kopparapu} R.~K.,  et~al., 2013a, \mn@doi [\apj]
  {10.1088/0004-637X/765/2/131}, \href
  {http://adsabs.harvard.edu/abs/2013ApJ...765..131K} {765, 131}

\bibitem[\protect\citeauthoryear{{Kopparapu} et~al.,}{{Kopparapu}
  et~al.}{2013b}]{Kopparapuerrata2013}
{Kopparapu} R.~K.,  et~al., 2013b, \mn@doi [\apj] {10.1088/0004-637X/770/1/82},
  \href {http://adsabs.harvard.edu/abs/2013ApJ...770...82K} {770, 82}

\bibitem[\protect\citeauthoryear{{Kopparapu}, {Ramirez}, {SchottelKotte},
  {Kasting}, {Domagal-Goldman}  \& {Eymet}}{{Kopparapu}
  et~al.}{2014}]{Kopparapu2014}
{Kopparapu} R.~K.,  {Ramirez} R.~M.,  {SchottelKotte} J.,  {Kasting} J.~F.,
  {Domagal-Goldman} S.,   {Eymet} V.,  2014, \mn@doi [\apjl]
  {10.1088/2041-8205/787/2/L29}, \href
  {http://adsabs.harvard.edu/abs/2014ApJ...787L..29K} {787, L29}

\bibitem[\protect\citeauthoryear{{Laakso}, {Rantala}  \&
  {Kaasalainen}}{{Laakso} et~al.}{2006}]{Laakso2006}
{Laakso} T.,  {Rantala} J.,   {Kaasalainen} M.,  2006, \mn@doi [\aap]
  {10.1051/0004-6361:20065121}, \href
  {http://adsabs.harvard.edu/abs/2006A%26A...456..373L} {456, 373}

\bibitem[\protect\citeauthoryear{{L\'ecuyer} \& {Gillet}}{{L\'ecuyer} \&
  {Gillet}}{1998}]{Lecuyer1998}
{L\'ecuyer} C.,  {Gillet} P.and~{Robert} F.,  1998, Chem. Geol., \href
  {http://adsabs.harvard.edu/abs/1998Icar..131..171K} {145, 249–261}

\bibitem[\protect\citeauthoryear{{Leinhardt} \& {Stewart}}{{Leinhardt} \&
  {Stewart}}{2012}]{LeinhardtStewart2012}
{Leinhardt} Z.~M.,  {Stewart} S.~T.,  2012, \mn@doi [\apj]
  {10.1088/0004-637X/745/1/79}, \href
  {http://adsabs.harvard.edu/abs/2012ApJ...745...79L} {745, 79}

\bibitem[\protect\citeauthoryear{{Lodders}, {Palme}  \& {Gail}}{{Lodders}
  et~al.}{2009}]{Lodders2009}
{Lodders} K.,  {Palme} H.,   {Gail} H.-P.,  2009, \mn@doi [Landolt
  B{\"o}rnstein] {10.1007/978-3-540-88055-4_34}, \href
  {http://adsabs.harvard.edu/abs/2009LanB...4B...44L} {}

\bibitem[\protect\citeauthoryear{{Mandell}, {Raymond}  \&
  {Sigurdsson}}{{Mandell} et~al.}{2007}]{Mandell2007}
{Mandell} A.~M.,  {Raymond} S.~N.,   {Sigurdsson} S.,  2007, \mn@doi [\apj]
  {10.1086/512759}, \href {http://adsabs.harvard.edu/abs/2007ApJ...660..823M}
  {660, 823}

\bibitem[\protect\citeauthoryear{{Marty}}{{Marty}}{2012}]{Marty2012}
{Marty} B.,  2012, \mn@doi [Earth and Planetary Science Letters]
  {10.1016/j.epsl.2011.10.040}, \href
  {http://adsabs.harvard.edu/abs/2012E$%$26PSL.313...56M} {313, 56}

\bibitem[\protect\citeauthoryear{{Marzari}}{{Marzari}}{2014}]{Marzari2014}
{Marzari} F.,  2014, \mn@doi [\mnras] {10.1093/mnras/stu929}, \href
  {http://adsabs.harvard.edu/abs/2014MNRAS.442.1110M} {442, 1110}

\bibitem[\protect\citeauthoryear{{Masset} \& {Snellgrove}}{{Masset} \&
  {Snellgrove}}{2001}]{MassetSnellgrove2001}
{Masset} F.,  {Snellgrove} M.,  2001, \mn@doi [\mnras]
  {10.1046/j.1365-8711.2001.04159.x}, \href
  {http://adsabs.harvard.edu/abs/2001MNRAS.320L..55M} {320, L55}

\bibitem[\protect\citeauthoryear{{Miguel}, {Guilera}  \& {Brunini}}{{Miguel}
  et~al.}{2011}]{Miguel2011}
{Miguel} Y.,  {Guilera} O.~M.,   {Brunini} A.,  2011, \mn@doi [\mnras]
  {10.1111/j.1365-2966.2011.19264.x}, \href
  {http://adsabs.harvard.edu/abs/2011MNRAS.417..314M} {417, 314}

\bibitem[\protect\citeauthoryear{{Morard}, {Bouchet}, {Valencia}, {Mazevet}  \&
  {Guyot}}{{Morard} et~al.}{2011}]{Morard2011}
{Morard} G.,  {Bouchet} J.,  {Valencia} D.,  {Mazevet} S.,   {Guyot} F.,  2011,
  \mn@doi [High Energy Density Physics] {10.1016/j.hedp.2011.02.001}, \href
  {http://adsabs.harvard.edu/abs/2011HEDP....7..141M} {7, 141}

\bibitem[\protect\citeauthoryear{{Morbidelli} \& {Crida}}{{Morbidelli} \&
  {Crida}}{2007}]{MorbidelliCrida2007}
{Morbidelli} A.,  {Crida} A.,  2007, \mn@doi [\icarus]
  {10.1016/j.icarus.2007.04.001}, \href
  {http://adsabs.harvard.edu/abs/2007Icar..191..158M} {191, 158}

\bibitem[\protect\citeauthoryear{{Mordasini}, {Alibert}, {Benz}  \&
  {Naef}}{{Mordasini} et~al.}{2009}]{Mordasini2009}
{Mordasini} C.,  {Alibert} Y.,  {Benz} W.,   {Naef} D.,  2009, \mn@doi [\aap]
  {10.1051/0004-6361/200810697}, \href
  {http://adsabs.harvard.edu/abs/2009A$%$26A...501.1161M} {501, 1161}

\bibitem[\protect\citeauthoryear{{Noack} et~al.,}{{Noack}
  et~al.}{2016}]{Noack2016}
{Noack} L.,  et~al., 2016, \mn@doi [\icarus] {10.1016/j.icarus.2016.05.009},
  \href {http://adsabs.harvard.edu/abs/2016Icar..277..215N} {277, 215}

\bibitem[\protect\citeauthoryear{{O'Brien}, {Morbidelli}  \&
  {Levison}}{{O'Brien} et~al.}{2006}]{OBrien2006}
{O'Brien} D.~P.,  {Morbidelli} A.,   {Levison} H.~F.,  2006, \mn@doi [\icarus]
  {10.1016/j.icarus.2006.04.005}, \href
  {http://adsabs.harvard.edu/abs/2006Icar..184...39O} {184, 39}

\bibitem[\protect\citeauthoryear{{Ohtsuki}, {Stewart}  \& {Ida}}{{Ohtsuki}
  et~al.}{2002}]{Ohtsuki2002}
{Ohtsuki} K.,  {Stewart} G.~R.,   {Ida} S.,  2002, \mn@doi [\icarus]
  {10.1006/icar.2001.6741}, \href
  {http://adsabs.harvard.edu/abs/2002Icar..155..436O} {155, 436}

\bibitem[\protect\citeauthoryear{{Paardekooper}, {Baruteau}, {Crida}  \&
  {Kley}}{{Paardekooper} et~al.}{2010}]{Paardekooper2010}
{Paardekooper} S.-J.,  {Baruteau} C.,  {Crida} A.,   {Kley} W.,  2010, \mn@doi
  [\mnras] {10.1111/j.1365-2966.2009.15782.x}, \href
  {http://adsabs.harvard.edu/abs/2010MNRAS.401.1950P} {401, 1950}

\bibitem[\protect\citeauthoryear{{Paardekooper}, {Baruteau}  \&
  {Kley}}{{Paardekooper} et~al.}{2011}]{Paardekooper2011}
{Paardekooper} S.-J.,  {Baruteau} C.,   {Kley} W.,  2011, \mn@doi [\mnras]
  {10.1111/j.1365-2966.2010.17442.x}, \href
  {http://adsabs.harvard.edu/abs/2011MNRAS.410..293P} {410, 293}

\bibitem[\protect\citeauthoryear{{Pfalzner}, {Steinhausen}  \&
  {Menten}}{{Pfalzner} et~al.}{2014}]{Pfalzner2014}
{Pfalzner} S.,  {Steinhausen} M.,   {Menten} K.,  2014, \mn@doi [\apjl]
  {10.1088/2041-8205/793/2/L34}, \href
  {http://adsabs.harvard.edu/abs/2014ApJ...793L..34P} {793, L34}

\bibitem[\protect\citeauthoryear{{Pringle}}{{Pringle}}{1981}]{Pringle1981}
{Pringle} J.~E.,  1981, \mn@doi [\araa] {10.1146/annurev.aa.19.090181.001033},
  \href {http://adsabs.harvard.edu/abs/1981ARA%26A..19..137P} {19, 137}

\bibitem[\protect\citeauthoryear{{Quintana}, {Barclay}, {Borucki}, {Rowe}  \&
  {Chambers}}{{Quintana} et~al.}{2016}]{Quintana2016}
{Quintana} E.~V.,  {Barclay} T.,  {Borucki} W.~J.,  {Rowe} J.~F.,   {Chambers}
  J.~E.,  2016, \mn@doi [\apj] {10.3847/0004-637X/821/2/126}, \href
  {http://adsabs.harvard.edu/abs/2016ApJ...821..126Q} {821, 126}

\bibitem[\protect\citeauthoryear{{Rafikov}}{{Rafikov}}{2004}]{Rafikov2004}
{Rafikov} R.~R.,  2004, \mn@doi [\aj] {10.1086/423216}, \href
  {http://adsabs.harvard.edu/abs/2004AJ....128.1348R} {128, 1348}

\bibitem[\protect\citeauthoryear{{Raymond} \& {Izidoro}}{{Raymond} \&
  {Izidoro}}{2017}]{RaymondIzidoro2017}
{Raymond} S.~N.,  {Izidoro} A.,  2017, \mn@doi [\icarus]
  {10.1016/j.icarus.2017.06.030}, \href
  {http://adsabs.harvard.edu/abs/2017Icar..297..134R} {297, 134}

\bibitem[\protect\citeauthoryear{{Raymond}, {Quinn}  \& {Lunine}}{{Raymond}
  et~al.}{2004}]{Raymond2004}
{Raymond} S.~N.,  {Quinn} T.,   {Lunine} J.~I.,  2004, \mn@doi [\icarus]
  {10.1016/j.icarus.2003.11.019}, \href
  {http://adsabs.harvard.edu/abs/2004Icar..168....1R} {168, 1}

\bibitem[\protect\citeauthoryear{{Raymond}, {Quinn}  \& {Lunine}}{{Raymond}
  et~al.}{2006}]{Raymond2006}
{Raymond} S.~N.,  {Quinn} T.,   {Lunine} J.~I.,  2006, \mn@doi [\icarus]
  {10.1016/j.icarus.2006.03.011}, \href
  {http://adsabs.harvard.edu/abs/2006Icar..183..265R} {183, 265}

\bibitem[\protect\citeauthoryear{{Raymond}, {O'Brien}, {Morbidelli}  \&
  {Kaib}}{{Raymond} et~al.}{2009}]{Raymond2009}
{Raymond} S.~N.,  {O'Brien} D.~P.,  {Morbidelli} A.,   {Kaib} N.~A.,  2009,
  \mn@doi [\icarus] {10.1016/j.icarus.2009.05.016}, \href
  {http://adsabs.harvard.edu/abs/2009Icar..203..644R} {203, 644}

\bibitem[\protect\citeauthoryear{{Raymond}, {Schlichting}, {Hersant}  \&
  {Selsis}}{{Raymond} et~al.}{2013}]{Raymond2013}
{Raymond} S.~N.,  {Schlichting} H.~E.,  {Hersant} F.,   {Selsis} F.,  2013,
  \mn@doi [\icarus] {10.1016/j.icarus.2013.06.019}, \href
  {http://adsabs.harvard.edu/abs/2013Icar..226..671R} {226, 671}

\bibitem[\protect\citeauthoryear{{Ricker} et~al.,}{{Ricker}
  et~al.}{2010}]{Ricker2010}
{Ricker} G.~R.,  et~al., 2010, in American Astronomical Society Meeting
  Abstracts \#215. p.~459

\bibitem[\protect\citeauthoryear{{Ronco} \& {de El{\'{\i}}a}}{{Ronco} \& {de
  El{\'{\i}}a}}{2014}]{Roncodeelia2014}
{Ronco} M.~P.,  {de El{\'{\i}}a} G.~C.,  2014, \mn@doi [\aap]
  {10.1051/0004-6361/201323313}, \href
  {http://adsabs.harvard.edu/abs/2014A%26A...567A..54R} {567, A54}

\bibitem[\protect\citeauthoryear{{Ronco}, {de El{\'{\i}}a}  \&
  {Guilera}}{{Ronco} et~al.}{2015}]{Ronco2015}
{Ronco} M.~P.,  {de El{\'{\i}}a} G.~C.,   {Guilera} O.~M.,  2015, \mn@doi
  [\aap] {10.1051/0004-6361/201526367}, \href
  {http://adsabs.harvard.edu/abs/2015A%26A...584A..47R} {584, A47}

\bibitem[\protect\citeauthoryear{{Ronco}, {Guilera}  \& {de
  El{\'{\i}}a}}{{Ronco} et~al.}{2017}]{Ronco2017}
{Ronco} M.~P.,  {Guilera} O.~M.,   {de El{\'{\i}}a} G.~C.,  2017, \mn@doi
  [\mnras] {10.1093/mnras/stx1746}, \href
  {http://adsabs.harvard.edu/abs/2017MNRAS.471.2753R} {471, 2753}

\bibitem[\protect\citeauthoryear{{Salvador Zain}, {de El{\'{\i}}a}, {Ronco}  \&
  {Guilera}}{{Salvador Zain} et~al.}{2017}]{Zain2017}
{Salvador Zain} P.,  {de El{\'{\i}}a} G.~C.,  {Ronco} M.~P.,   {Guilera} O.~M.,
   2017, preprint, \href {http://adsabs.harvard.edu/abs/2017arXiv171004617S} {}
  (\mn@eprint {arXiv} {1710.04617})

\bibitem[\protect\citeauthoryear{{Schneider}, {Dedieu}, {Le Sidaner}, {Savalle}
   \& {Zolotukhin}}{{Schneider} et~al.}{2011}]{Schneider2011}
{Schneider} J.,  {Dedieu} C.,  {Le Sidaner} P.,  {Savalle} R.,   {Zolotukhin}
  I.,  2011, \mn@doi [\aap] {10.1051/0004-6361/201116713}, \href
  {http://adsabs.harvard.edu/abs/2011A%26A...532A..79S} {532, A79}

\bibitem[\protect\citeauthoryear{{Selsis}, {Kasting}, {Levrard}, {Paillet},
  {Ribas}  \& {Delfosse}}{{Selsis} et~al.}{2007}]{Selsis2007}
{Selsis} F.,  {Kasting} J.~F.,  {Levrard} B.,  {Paillet} J.,  {Ribas} I.,
  {Delfosse} X.,  2007, \mn@doi [\aap] {10.1051/0004-6361:20078091}, \href
  {http://adsabs.harvard.edu/abs/2007A%26A...476.1373S} {476, 1373}

\bibitem[\protect\citeauthoryear{{Simpson}}{{Simpson}}{2017}]{Simpson2017}
{Simpson} F.,  2017, \mn@doi [\mnras] {10.1093/mnras/stx516}, \href
  {http://adsabs.harvard.edu/abs/2017MNRAS.468.2803S} {468, 2803}

\bibitem[\protect\citeauthoryear{{Suzuki}, {Muto}  \& {Inutsuka}}{{Suzuki}
  et~al.}{2010}]{Suzuki2010}
{Suzuki} T.~K.,  {Muto} T.,   {Inutsuka} S.-i.,  2010, \mn@doi [\apj]
  {10.1088/0004-637X/718/2/1289}, \href
  {http://adsabs.harvard.edu/abs/2010ApJ...718.1289S} {718, 1289}

\bibitem[\protect\citeauthoryear{{Suzuki}, {Ogihara}, {Morbidelli}, {Crida}  \&
  {Guillot}}{{Suzuki} et~al.}{2016}]{Suzuki2016}
{Suzuki} T.~K.,  {Ogihara} M.,  {Morbidelli} A.,  {Crida} A.,   {Guillot} T.,
  2016, \mn@doi [\aap] {10.1051/0004-6361/201628955}, \href
  {http://adsabs.harvard.edu/abs/2016A%26A...596A..74S} {596, A74}

\bibitem[\protect\citeauthoryear{{Tanaka}, {Takeuchi}  \& {Ward}}{{Tanaka}
  et~al.}{2002}]{Tanaka2002}
{Tanaka} H.,  {Takeuchi} T.,   {Ward} W.~R.,  2002, \mn@doi [\apj]
  {10.1086/324713}, \href {http://adsabs.harvard.edu/abs/2002ApJ...565.1257T}
  {565, 1257}

\bibitem[\protect\citeauthoryear{{Tanigawa} \& {Ikoma}}{{Tanigawa} \&
  {Ikoma}}{2007}]{TanigawaIkoma2007}
{Tanigawa} T.,  {Ikoma} M.,  2007, \mn@doi [\apj] {10.1086/520499}, \href
  {http://adsabs.harvard.edu/abs/2007ApJ...667..557T} {667, 557}

\bibitem[\protect\citeauthoryear{{Touboul}, {Kleine}, {Bourdon}, {Palme}  \&
  {Wieler}}{{Touboul} et~al.}{2007}]{Touboul2007}
{Touboul} M.,  {Kleine} T.,  {Bourdon} B.,  {Palme} H.,   {Wieler} R.,  2007,
  \mn@doi [\nat] {10.1038/nature06428}, \href
  {http://adsabs.harvard.edu/abs/2007Natur.450.1206T} {450, 1206}

\bibitem[\protect\citeauthoryear{{Tsiganis}, {Gomes}, {Morbidelli}  \&
  {Levison}}{{Tsiganis} et~al.}{2005}]{Tsiganis2005}
{Tsiganis} K.,  {Gomes} R.,  {Morbidelli} A.,   {Levison} H.~F.,  2005, \mn@doi
  [\nat] {10.1038/nature03539}, \href
  {http://adsabs.harvard.edu/abs/2005Natur.435..459T} {435, 459}

\bibitem[\protect\citeauthoryear{{Valencia}, {O'Connell}  \&
  {Sasselov}}{{Valencia} et~al.}{2007}]{Valencia2007}
{Valencia} D.,  {O'Connell} R.~J.,   {Sasselov} D.~D.,  2007, \mn@doi [\apjl]
  {10.1086/524012}, \href {http://adsabs.harvard.edu/abs/2007ApJ...670L..45V}
  {670, L45}

\bibitem[\protect\citeauthoryear{{Veras} \& {Evans}}{{Veras} \&
  {Evans}}{2013}]{VerasEvans2013}
{Veras} D.,  {Evans} N.~W.,  2013, \mn@doi [\mnras] {10.1093/mnras/sts647},
  \href {http://adsabs.harvard.edu/abs/2013MNRAS.430..403V} {430, 403}

\bibitem[\protect\citeauthoryear{{Walsh}, {Morbidelli}, {Raymond}, {O'Brien}
  \& {Mandell}}{{Walsh} et~al.}{2011}]{Walsh2011}
{Walsh} K.~J.,  {Morbidelli} A.,  {Raymond} S.~N.,  {O'Brien} D.~P.,
  {Mandell} A.~M.,  2011, \mn@doi [\nat] {10.1038/nature10201}, \href
  {http://adsabs.harvard.edu/abs/2011Natur.475..206W} {475, 206}

\bibitem[\protect\citeauthoryear{{Williams} \& {Pollard}}{{Williams} \&
  {Pollard}}{2002}]{WilliamsPollard2002}
{Williams} D.~M.,  {Pollard} D.,  2002, \mn@doi [International Journal of
  Astrobiology] {10.1017/S1473550402001064}, \href
  {http://adsabs.harvard.edu/abs/2002IJAsB...1...61W} {1, 61}

\bibitem[\protect\citeauthoryear{{Wolfgang}, {Rogers}  \& {Ford}}{{Wolfgang}
  et~al.}{2016}]{Wolfgang2016}
{Wolfgang} A.,  {Rogers} L.~A.,   {Ford} E.~B.,  2016, \mn@doi [\apj]
  {10.3847/0004-637X/825/1/19}, \href
  {http://adsabs.harvard.edu/abs/2016ApJ...825...19W} {825, 19}

\bibitem[\protect\citeauthoryear{{de El{\'{\i}}a}, {Guilera}  \& {Brunini}}{{de
  El{\'{\i}}a} et~al.}{2013}]{deElia2013}
{de El{\'{\i}}a} G.~C.,  {Guilera} O.~M.,   {Brunini} A.,  2013, \mn@doi [\aap]
  {10.1051/0004-6361/201321304}, \href
  {http://adsabs.harvard.edu/abs/2013A$%$26A...557A..42D} {557, A42}

\makeatother
\end{thebibliography}









\bsp	
\label{lastpage}
\end{document}